\documentclass[aps,prb,twocolumn,superscriptaddress,preprintnumbers,10pt]{revtex4-2}
\usepackage[colorlinks, allcolors=Blue, linktocpage]{hyperref}
\usepackage[dvipsnames]{xcolor}
\usepackage{amsmath,amssymb,graphicx}
\usepackage{newtxtext,newtxmath,mathrsfs,multirow,enumitem}
\usepackage[boxsize=.5em, aligntableaux=center]{ytableau}
\usepackage{comment}
\usepackage{booktabs}
\setlist{align=parleft,leftmargin=0pt,itemindent=2.5em,labelsep=0em,labelwidth=1.5em,nosep}

\usepackage{dsfont}
\newcommand{\one}{\mathbb{I}}
\newcommand{\Dh}{\hat\Delta}
\newcommand{\Norb}{N_m}
\newsavebox{\tabbox}
\newcommand{\fitwidth}[1]{\sbox{\tabbox}{#1}\ifdim\wd\tabbox>\textwidth\resizebox{\textwidth}{!}{\usebox{\tabbox}}\else\usebox{\tabbox}\fi}

\usepackage[normalem]{ulem}
\definecolor{orange}{rgb}{1,0.5,0}
\definecolor{darkblue}{rgb}{0.,0.,0.6}
\definecolor{darkred}{rgb}{0.5,0.,0.}

\newcommand{\delete}[1]{ {\color{red}  \sout{#1}   }}

\begin{document}

\preprint{KYUSHU-HET-378}

\title{Monodromy Defects of 3D Wilson--Fisher Criticality on the Fuzzy Sphere }

\author{Yin Tang}\email[]{tang.yin.129@m.kyushu-u.ac.jp}
\affiliation{Institute for Advanced Study, Kyushu University, Fukuoka 819-0395, Japan}

\author{W. Zhu}\email[]{zhuwei@westlake.edu.cn}
\affiliation{Westlake Institute of Advanced Study, Westlake University, Hangzhou 310024, China}

\begin{abstract}
Monodromy (twist) line defects arise as the boundaries of topological symmetry surfaces, which  provide a natural route to  conformal defects with  profound  implications in condensed matter, gauge theories and quantum information theory. However, their interacting dynamics remains difficult to characterize non-perturbatively.
Here we design a way to realize the monodromy
boundary conditions in the fuzzy-sphere regularization by threading a symmetry-selective Aharonov--Bohm flux through the two poles of the sphere. 
We apply this construction to the $\mathbb{Z}_2$ twist of the 3D Ising conformal field theory (CFT), $\pi$-twist of the 3D O(2) CFT and three inequivalent monodromies of O(3) (centre twist, z-axis $\pi$-twist and the reflection). For each defect,  
we extract the
low-lying defect spectrum together with bulk-to-defect and defect-defect operator product expansion coefficients, which characterize the operators living on the defect line, their mutual interactions, and how they couple to the surrounding critical bulk. 
Among the low-lying defect primaries resolved in our spectra, we find no relevant scalar perturbation that preserves all the symmetries of the corresponding defect. The relevant deformations identified in our calculations break at least one preserved symmetry, providing evidence that these monodromy defects are stable against symmetry-preserving relevant perturbations. For the partial twists of O(3),  %
we derive the flow of the symmetry-preserving  %
untwisted zero-mode mass  generated by the bulk interaction. 
Its weak-coupling runaway near four dimensions motivates a careful distinction between the microscopic absence of pinning, the free reference background, and a possible infrared defect fixed point in three dimensions. These results establish a systematic set of non-perturbative conformal data for monodromy defects beyond the Ising twist. Our construction therefore turns the symmetry flux into a tunable knob of the fuzzy-sphere regularization, opening a route to exploring the response of critical matter to distinct defects.

\end{abstract}

\maketitle

\section{Introduction}

Physical observables are commonly connected to local operators which operate at individual spacetime points. However, the observables of a quantum field theory can be extended operators, supported on lines, surfaces, or higher-dimensional submanifolds. Such operators describe physical structures ranging from the trajectories of external probes to interfaces between distinct phases, and expose aspects of a theory that are not readily apparent from its local observables. A familiar example is the Wilson loop operator  \cite{Wilsonloop}, which diagnoses confinement through its large-loop behavior and plays  a fundamental role in topological quantum field theory \cite{Witten1989,Fradkin_book}.
In particular, extended operators that preserve the conformal symmetry ~\cite{Polyakov:1970xd,Cardy:1996xt,DiFrancesco:1997nk} compatible with their geometry define conformal defects~\cite{Billo:2016cpy,Liendo:2012hy,Andrei:2018die}.  These defects support their own universal data and govern the critical behaviors in the presence of impurities, boundaries, and interfaces. 
Extended operators therefore connect the local description of critical phenomena to the physics of probes~\cite{Assaad:2013xua,ParisenToldin:2016vre,Cuomo:2021kfm}, defects ~\cite{Sachdev:1999bk,Vojta:2000qi,Allais:2014fqa,Cuomo:2022xgw}, renormalization group flow ~\cite{Cuomo:2021rkm} and topological organization~\cite{Aharony:2022ntz,Aharony:2023amq,Krishnan:2023cff}, making them essential ingredients in a broader understanding of conformal field theories (CFTs).

\begin{figure}[t]
\centering
\includegraphics[width=0.9\linewidth]{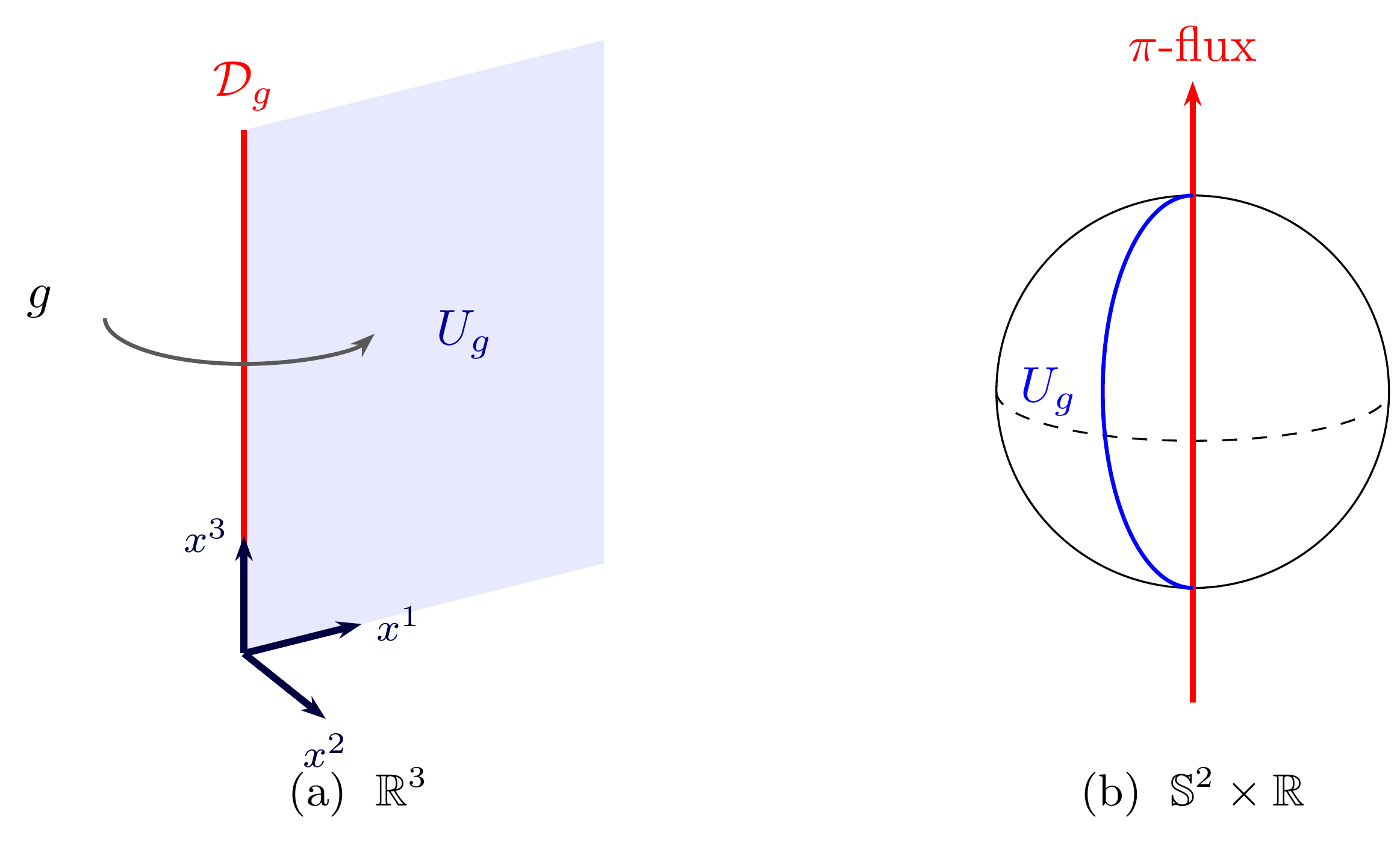}
\caption{(a) A monodromy defect $\mathcal{D}_g$ is the boundary of the topological surface $U_g$. 
(b) In radial quantization a straight defect through the origin pierces each time slice $S^2$ at the two poles, and $U_g$ intersects it along a meridian, across which the $g$-odd fields change sign. 
Twisting a $g$-odd fermion is realized by an Aharonov--Bohm $\pi$-flux through the poles of two-sphere, which shifts the orbitals of these fermions to half-integer $m$ (Sec.~\ref{sec:construction}).
}
\label{fig:cartoon}
\end{figure}

In this work we consider a class of line defects with profound physics: \emph{monodromy} (or twist) defects~\cite{Kadanoff:1970kz,Gukov:2006jk,Billo:2013jda,Gaiotto:2013nva}.
As shown in Fig. \ref{fig:cartoon}(a), given a global symmetry element $g$, the topological surface operator $U_g$ ends on a line $\mathcal D_g$; fields charged under $g$ then pick up the monodromy $g$ when transported around the line $\mathcal D_g$.
In the perspective of generalized global symmetries, a monodromy defect is the boundary of a topological
symmetry operator of codimension one \cite{Gaiotto:2014kfa}.
Monodromy defects provide a setting in which symmetry and topology constrain an interacting system, which has wide applications in diverse areas of physics. 
For example, a monodromy line is a flux tube of the background gauge field that couples to the symmetry, so its universal data describe how a critical system responds to the insertion of symmetry flux, such as a vortex line in a superfluid or a spin-flux tube in a quantum magnet~\cite{Komargodski:2025spf}. 
A charge transported around such a flux acquires an Aharonov--Bohm phase, so the operators living on the line carry fractional transverse spin~\cite{Wilczek:1982wy}; the defect spectrum is thus organized by a quantum number that has no counterpart on the trivial line.
Under particle--vortex duality~\cite{Peskin:1977kp,Dasgupta:1981zz,Karch:2016sxi,Seiberg:2016gmd} the U(1) monodromy lines of the O(2) CFT become Wilson lines of fractional charge in the critical Abelian Higgs model. 
Monodromy defects also implement the cyclic permutation of replicas in the replica approach to Renyi entropies~\cite{Hung:2014npa,Bianchi:2015liz}.

Technically,   a monodromy defect is challenging  to implement because fields must obey a symmetry-twisted boundary condition across a branch surface, with a consistent regularization of its boundary.
Despite progress using perturbative methods such as the $\epsilon$- and large-$N$
expansions~\cite{Yamaguchi:2016pbj,Soderberg:2017oaa,Giombi:2021uae,Bianchi:2021snj} and the large-transverse-spin
expansion~\cite{Lemos:2017vnx}, non-perturbative tools like numerical methods are highly desired. Previous attempts have been made in the Ising criticality 
by Monte Carlo simulation~\cite{Billo:2013jda}, and the Ising and O(2) CFTs by the bootstrap~\cite{Gaiotto:2013nva,GimenezGrau:2021wiv,GimenezGrau:2022czc}. 
Both approaches have limitations. Monte Carlo simulations do not resolve the conformal multiplet structure directly and require a dedicated and costly measurement for each observable, so that only a handful of defect data are accessible. 
The defect bootstrap faces a structural difficulty: positivity is lost in the bulk channel of bulk two-point functions, so that existing studies rely on additional spectral assumptions~\cite{Gaiotto:2013nva,GimenezGrau:2022czc}. Thus, a non-perturbative method that yields the defect spectrum together with conformal data, directly and at modest cost, is desirable.

Recently, the fuzzy-sphere regularization~\cite{Zhu:2022gjc,He:2026fsj} has emerged as a powerful non-perturbative tool for
3D CFTs. Interacting fermions in the lowest Landau level of a sphere surrounding a magnetic
monopole~\cite{Haldane:1983xm,Wu:1976ge} realize a non-commutative (fuzzy) sphere~\cite{Madore:1991bw} on which the
CFT lives directly on the cylinder $S^2\times\mathbb{R}$, so that the state-operator correspondence~\cite{Cardy:1984rp,Cardy:1985lth} gives the operator spectrum directly. 
The fuzzy-sphere scheme can access not only the scaling dimensions of CFT operators~\cite{Zhu:2022gjc},
but also the operator product expansion (OPE) coefficients~\cite{Hu:2023xak}, multi-point correlators~\cite{Han:2023yyb}, the entropic
$F$-function~\cite{Hu:2024pen}, $g$-function ~\cite{Zhou:2023fqu}, conformal generators~\cite{Fardelli:2024qla,Fan:2024vcz}, and conformal defects ~\cite{Hu:2023ghk,Zhou:2023fqu,Cuomo:2024psk}.
This scheme  has been applied to  a range of strongly interacting CFTs, including O($N$) Wilson--Fisher CFTs~\cite{Han:2023lky,Guo:2025onf,Dey:2025o3,Dey:2026o2,Stergiou:2026cub}, Lee-Yang non-unitary CFT \cite{ArguelloCruz:2025zuq,Fan:2025bhc,EliasMiro:2025msj}, deconfined
criticality~\cite{Zhou:2023qfi,Zhou:2024spn,Yang:2026cof,Chen:2023xjc}, free scalar CFT \cite{He:2025ong,Taylor:2025odf}, topological phase transitions~\cite{Voinea:2026maj,Zhou:2025fer,Gao:2025gny,Zhou:2025csm}, 
superconformal CFTs \cite{Zhou:2025fer,Tang:2026sci}, surface CFTs~\cite{Zhou:2024dbt,Dedushenko:2024nwi,Feng:2026sur}, defect CFTs~\cite{Hu:2023ghk,Zhou:2023fqu,Yang:2026onl,Liu:2026dqc}, and spin impurities~\cite{Sarma:2026bk}.

Here we show that the monodromy boundary condition can be built into the fuzzy sphere exactly, by threading a symmetry-selective Aharonov--Bohm $\pi$-flux through the two poles: the flavour symmetry-odd fermions are placed on lowest-Landau-level orbitals whose angular-momentum grid is shifted by half a unit. The construction preserves the symmetries of the defect exactly and applies to any involution acting diagonally on the fermion flavours. As a showcase application, we use it to obtain conformal data (the defect spectra, bulk-to-defect and defect-defect OPE coefficients) of five distinct monodromy defects: the $\mathbb Z_2$ twist of the Ising CFT, the $\pi$-monodromy of the O(2) CFT and three inequivalent monodromies of the O(3) CFT.

Three physical lessons are learned from our results. First, the Aharonov--Bohm organization of the spectrum survives at the strongly coupled fixed point, evidenced by the emergence of conformal defects with defect operators carrying fractional transverse spins.  Second, a systematic comparison with perturbation theory shows that  many low-lying twisted modes and their composites  are close to  a generalized-free-field description. Their leading interaction energies have the  sign and flavour hierarchy of the one-loop  contact term, with strengths typically enhanced by factors of $1.1$--$1.8$  in the channels compared below.
Third, no symmetry-preserving relevant scalar is resolved in the spectra of the five defects, providing evidence for stability of the conformal interpretation at the sizes studied. The resolved relevant operators instead generate symmetry-breaking  deformations, including monodromy pinning~\cite{Kravchuk:2025mpd}.

The paper is organized as follows. In Sec.~\ref{sec:theory}, we introduce monodromy defects regarding distinct symmetries in 3D Wilson--Fisher CFTs. We then apply a one-loop analysis in $d=4-\epsilon$ for arbitrary partial
twists of the O($M$) model (Sec. \ref{sec:perturbation}), and comment on the zero-mode defect mass of the partial twists (App.~\ref{sec:untwtheory}).   Following that, in Sec.~\ref{sec:fuzzy},
we explain how to realize the monodromy (twist) boundary conditions in the fuzzy sphere.
We then present our numerical results in Sec.~\ref{sec:results}. We provide strong evidence for the formation of conformal defects in
five different universality classes and present their conformal data. Finally, we present a summary and discussion in Sec.~\ref{sec:summary}.  Complementary  data are in the appendices.


\section{Monodromy defects in O($N$) CFTs}
\label{sec:theory}

\subsection{Definition and symmetries}
\label{sec:def}
Let $U_g[\Sigma]$ be the corresponding topological surface operator with the symmetry element $g\in O(N)$. If $\Sigma$ has a
boundary $\partial\Sigma=\gamma$, the configuration defines a line defect $\mathcal{D}_g[\gamma]$ (see Fig. \ref{fig:cartoon}(a)), where the surface
itself is invisible and only its boundary is physical. For a straight line along the $\tau$ direction with
transverse polar coordinates $(r,\varphi)$, a field $\phi_a$ obeys 
\begin{equation}
\phi_a(r,\varphi+2\pi,\tau)=g_{ab}\,\phi_b(r,\varphi,\tau).
\end{equation}
The monodromy preserves the subgroup of O($N$) commuting with
$g$ and transverse rotations $SO(2)_\perp$ generated by $L_z$.  At a conformal defect fixed point it also preserves the $SL(2,\mathbb{R})$ conformal symmetry along the line. In what follows, we consider $\mathbb Z_2$ monodromies ($g^2=\one$) in O($N$) Wilson-Fisher CFTs (see Tab. \ref{tab:defects}):
\begin{itemize}
\item Ising ($N=1$): the $\mathbb{Z}_2$ twist $g=-1$, preserving $\mathbb{Z}_2$;
\item O(2): the $\pi$-rotation $g=R(\pi)=-\one$, preserving the full O(2);
\item O(3) centre twist: $g=-\one$, preserving the full O(3);
\item O(3) $R_z(\pi)$: $g=\mathrm{diag}(-1,-1,+1)$, preserving O(2)$\times\mathbb{Z}_2$, with one \emph{untwisted}
component $\phi_3$;
\item O(3) reflection: $g=\mathrm{diag}(+1,+1,-1)$, preserving O(2)$\times\mathbb{Z}_2$, with only $\phi_3$ twisted and an
\emph{untwisted O(2) doublet} $(\phi_1,\phi_2)$.
\end{itemize}

The three O(3) defects are inequivalent. Since $-\one$ is central and has determinant $-1$ in three dimensions,
$O(3)=SO(3)\times\{\pm\one\}$, and the three elements are $(\one,-\one)$, $(R_z(\pi),+\one)$ and $(R_z(\pi),-\one)$: they
belong to different conjugacy classes, distinguished by the determinant and by the number $n_{\rm tw}=3,2,1$ of
eigenvalues $-1$. In particular the reflection is the product of the centre twist and $R_z(\pi)$, so that the corresponding group-like
topological surface operators obey the fusion rule $U_{\rm refl}=U_{-\one}U_{R_z(\pi)}$; the three line defects are nevertheless
distinct, as their spectra confirm (Sec.~\ref{sec:o3}).

\subsection{Defect operators and observables}
Defect primaries $\hat{\mathcal{O}}_s$ carry a transverse spin $s$ under $SO(2)_\perp$, which is
half-integer for operators odd under $g$ and integer otherwise. (Hereafter, we denote the defect operator with a hat $\hat{\mathcal{O}}$, while the bulk operator as  $\mathcal{O}$ without a hat.) The bulk-to-defect expansion of a component of the order parameter with monodromy $g_a=\pm1$ reads
\begin{equation}
\phi_a(x)\sim \sum_s b_{\phi\psi_s}\,\frac{e^{is\varphi}}{r^{\Delta_\phi-\Dh_s}}\,\hat{\psi}_{a,s}(\tau),\quad s\in\begin{cases}\mathbb{Z}+\tfrac12 & g_a=-1\\ \mathbb Z & g_a=+1\end{cases}
\label{eq:bdexp}
\end{equation}
where $\Dh_s$ is the scaling dimension of the defect operator $\hat{\psi}_s$ and $b_{\phi\hat{\psi}_s}$ is the bulk-to-defect OPE.  
In addition, every conformal line defect contains the displacement operator $\hat D$ with $s=\pm1$ and
$\Dh_{\hat D}=2$, which we use as an internal check and as an alternative normalization of the spectrum. Scalar bulk
operators invariant under the preserved symmetry acquire one-point functions
$\langle\mathcal{O}(x)\rangle_{\mathcal D}=a_{\mathcal{O}}/r^{\Delta_{\mathcal O}}$. The Ward identity of the displacement
operator relates $a_{\mathcal O}$ to the bulk-to-defect coupling $b_{\mathcal O\hat D}$ and to the normalization $C_D$ of the
displacement two-point function~\cite{Billo:2016cpy}. We use this
relation only as a consistency check and discuss it together with the O(3) one-point functions in App.~\ref{app:O3}.

\subsection{One-loop analysis}
\label{sec:perturbation}
For a twist of all components ($n_{\rm tw}=M$) the one-loop dimensions of the single-mode operators are computed for Ising~\cite{Gaiotto:2013nva,Yamaguchi:2016pbj} and O($M$)~\cite{Soderberg:2017oaa,Giombi:2021uae}, and those of a few two-mode operators of the Ising twist defect were obtained in Ref.~\cite{Gaiotto:2013nva}. Here we need the generalization to partial twists, which contain untwisted components, and the  contact matrix elements of general two-mode states, including the need for degenerate perturbation theory. We derive these quantities in App.~\ref{sec:oneloop} and summarize the results here.
The quartic coupling has two effects at one loop. First, the tadpole of the twisted fields induces an attractive potential $\mu_a\phi_a^2/(2r^2)$ near the line, stronger for twisted than for untwisted components,
\begin{equation}\label{eq:eqmu}
\mu_{\rm tw}=-\frac{(n_{\rm tw}+2)\,\epsilon}{4(M+8)},\qquad \mu_{\rm untw}=-\frac{n_{\rm tw}\,\epsilon}{4(M+8)},
\end{equation}
which gives, for $s\neq0$ and $|\mu|\ll s^2$, the single-mode shift $\Dh_s=1-\frac{\epsilon}{2}+|s|+\frac{\mu}{2|s|}+O(\epsilon^2)$ on this reference background. Equation~\eqref{eq:eqmu} is the self-energy evaluated on the free (reference) monodromy background, in which every untwisted $s=0$ mode keeps its free boundary condition. For a full twist, perturbation theory about the regular free defect gives the fixed-point data to this order. For a partial twist the untwisted $s=0$ channel is exceptional: the expansion in $\mu/|s|$ breaks down, the defect operator $\hat\phi_0^2$ is classically marginal, and the bulk interaction generates the symmetry-preserving defect mass $\frac{c}{2}\int\hat\phi_0^2$ even when it is absent microscopically. This flow was analysed in Ref.~\cite{MCheng2026a}. In this work we keep the free boundary condition of the untwisted zero mode in all one-loop reference values (App.~\ref{sec:untwtheory}). Second, a two-mode operator $[\psi_{s_1}\psi_{s_2}]$ receives a contact term $\gamma^{\rm c}=\frac{3\epsilon}{M+8}\,T_R\,\frac{2}{1+|s_1|+|s_2|}$, halved for identical modes, whose flavour weight $T_R$ is $\frac{M'+2}{3}$ for the singlet of the symmetry O($M'$) acting on the pair, $\frac{2}{3}$ for the symmetric traceless and $0$ for the antisymmetric representation, and $\frac{1}{3}$ for a twisted--untwisted pair. For a full twist, the singlet of two identical $\psi_{1/2}$ modes is the displacement operator and the one-loop terms cancel to give $\Dh_{\hat D}=d-1$. Partial twists additionally require the symmetry-allowed mixing discussed in App.~\ref{sec:oneloop}.


\section{Fuzzy-sphere realization}
\label{sec:fuzzy}

We consider fermions living on a two-sphere with monopole charge $s$~\cite{Haldane:1983xm}.
The induced lowest Landau level hosts $\Norb=2s+1$ degenerate Landau orbitals. Accordingly, 
a fermion local field in real-space is expanded by
$\psi_\alpha(\mathbf{x})=R^{-1}\sum_m Y^{(s)}_m(\hat{\mathbf n})c_{m,\alpha}$, where $c_{m,\alpha}$ annihilates a fermion on a Landau orbital $m =-s,\dots,s$, $R=\sqrt{\Norb}$ is the radius of the sphere and $Y^{(s)}_m(\theta,\varphi)$ is the monopole harmonics \cite{Wu:1976ge}
\begin{equation}
Y^{(s)}_m(\theta,\varphi)\propto e^{im\varphi}\cos^{s+m}\tfrac{\theta}{2}\,\sin^{s-m}\tfrac{\theta}{2}.
\label{eq:orb}
\end{equation}

In particular, in the simulation of a magnetic phase transition, all the gapless degrees of freedom are charge-neutral which should be bilinears of fermions or combinations of even number of fermions. For example, the order parameter of the Ising transition is a particle-hole excitation of fermions \cite{Zhu:2022gjc},
\begin{equation} \label{eq:orderp}
\sigma \sim n^z = \psi^\dagger_\alpha\sigma^z_{\alpha\beta} \psi_\beta 
\end{equation}
This identification allows us to implement monodromy of the critical order parameter through the fermion orbitals.

\begin{table*}[t]
\centering\footnotesize
\caption{The monodromy defects considered in this work. Flavour bases are described in Sec.~\ref{sec:bases} and 
``twisted'' flavours use the half-shifted orbital grid~\eqref{eq:twisted}. 
The single-particle (sp.) count is
given at $\Norb=5$ for example.}
\label{tab:defects}
\fitwidth{%
\begin{tabular}{llllll}
\toprule
CFT & $g$ & twisted flavours & sp.\ states ($\Norb=5$) & preserved symmetry & sector labels \\
\midrule
Ising & $\mathbb{Z}_2$ & $\psi_-$ & $5+4=9$ & $\mathbb{Z}_2$ & $2L_z$ ($\mathbb{Z}_2$ charge $=2L_z$ mod 2) \\
O(2) & $R(\pi)=-\one$ & $\psi_+,\psi_-$ & $4+5+4=13$ & O(2) & $Q$, $2L_z$ ($2L_z\equiv Q$), $C$ at $Q=0$ \\
O(3) & $-\one$ & $\psi_+,\psi_z,\psi_-$ & $4+4+4+5=17$ & O(3) & $S^z$, $2L_z$, $S$ \\
O(3) & $R_z(\pi)$ & $\psi_+,\psi_-$ & $4+5+4+5=18$ & O(2)$\times\mathbb{Z}_2$ & $S^z$, $2L_z$ ($2L_z\equiv S^z$), $P_3$ \\
O(3) & $\mathrm{diag}(1,1,-1)$ & $\psi_z$ & $5+4+5+5=19$ & O(2)$\times\mathbb{Z}_2$ & $S^z$, $2L_z$ ($g$-charge $=2L_z$ mod 2), $C$ at $S^z=0$ \\
\bottomrule
\end{tabular}}
\end{table*}

\subsection{Monodromy defects from half-shifted orbitals}
\label{sec:construction}

In radial quantization a straight defect through the origin intersects every
time slice $S^2$ at the north and south poles, and the topological symmetry surface intersects the time slice
along a meridian from pole to pole (Fig.~\ref{fig:cartoon}(b)). Fermion flavours odd under $g$ therefore acquire a
sign when transported once around the $z$-axis relative to the untwisted flavours. The untwisted flavours have
$m=-s,-s+1,\dots,s$ 
and the twist is implemented by shifting the angular-momentum grid of the $g$-odd flavours by half a unit,
\begin{equation}
m=-s+\tfrac12,\,-s+\tfrac32,\,\dots,\,s-\tfrac12 ,
\label{eq:twisted}
\end{equation}
$2s$ orbitals in total. For odd $\Norb$ these are the half-integer-$m$ orbitals; for even $\Norb$ they have integer $m$. For example, at $\Norb=10$ ($s=\frac{9}{2}$) an untwisted flavour has $m\in\{-\frac{9}{2},\dots,\frac{9}{2}\}$ and a
twisted one $m\in\{-4,\dots,4\}$. In all cases the exponents $s\pm m$ of the twisted orbitals~\eqref{eq:orb} are
half-integers $\ge\frac{1}{2}$, so the orbitals are regular at both poles, where they vanish at least as fast as $\theta^{1/2}$ and
$(\pi-\theta)^{1/2}$, and every bilinear with one twisted and one untwisted leg carries half-integer $L_z$, i.e.\ is antiperiodic
around the line. Equivalently, a singular gauge transformation $\psi\to e^{\pm i\varphi/2}\psi$ of the twisted flavours maps them
to the untwisted grid in the presence of an Aharonov–
Bohm flux $\Phi_0/2$ along the $z$-axis; the regularity condition at the poles removes one orbital.

We make several remarks in order. 
(i) Every $g$-odd fermion bilinear, e.g.\ $\sigma\sim n^z$ in Ising (Eq. \ref{eq:orderp}), acquires half-integer $L_z$, so the Hilbert space decomposes into sectors with
$L_z\in\mathbb{Z}+q/2$, where $q$ is the $g$-charge. These are exactly the transverse spins of the defect operators.
This provides a natural implementation of monodromy boundary conditions in the fuzzy-sphere regularization.
(ii)  The real-space Hamiltonian density and all bulk couplings are kept fixed; we change the orbital sets to impose the monodromy and recompute the projected matrix elements. No independently adjustable one-body or interaction term is added at either pole. (iii) The  core regions have a width of order the magnetic length and determine the short-distance matching to continuum defect couplings. For a partial twist, this includes the symmetry-preserving mass of the untwisted zero mode. Thus ``unpinned'' refers to the absence of an explicit microscopic pole term, not to a renormalized mass constrained to vanish at every scale. Ref.~\cite{MCheng2026a} studies both zero and nonzero explicit lattice mass pinning; our construction fixes the former type of microscopic prescription throughout. Different regularizations need not assign the same renormalized initial coupling to zero explicit pinning. Whether they flow to the same infrared defect is a question of the  basin of attraction, not of monodromy alone.
(iv) The construction preserves $L_z$,
the commutant of $g$ in O($N$), and the pole-exchanging rotation $\mathcal R_y(\pi)$.

\subsection{Flavour bases and twisted orbitals, model by model}
\label{sec:bases}

The construction of Sec.~\ref{sec:construction} only changes the single-particle
orbital set of a flavour, so it applies when $g$ acts \emph{diagonally} on the fermion flavours with
eigenvalues $\pm1$. We therefore first rotate to a flavour basis $\{\psi_\alpha\}$ in which
$g=\mathrm{diag}(g_\alpha)$, $g_\alpha=\pm1$, and then assign to each flavour the orbital set
\begin{equation}
\mathcal M_\alpha=\begin{cases}\{-s,-s+1,\dots,s\} & g_\alpha=+1,\\[2pt]
\{-s+\tfrac12,-s+\tfrac32,\dots,s-\tfrac12\} & g_\alpha=-1,\end{cases}
\label{eq:Malpha}
\end{equation}
with $2s=\Norb-1$. The Fock space is spanned by the occupation numbers of the single-particle states
$(\alpha,m)$, $m\in\mathcal M_\alpha$, at fixed electron number $N_e=\Norb$. Once $\mathcal M_\alpha$ changes, all projected matrix elements are recomputed using the same bulk interaction parameters ($V_0,V_1$ for Ising, and $U_1$ for O($N$); App.~\ref{app:me}). For odd $N_m$, twisted orbitals carry odd $2m$ and untwisted ones even $2m$, and vice versa for even $N_m$; in either case, since $N_e=N_m$, we have
\begin{equation}
2L_z\equiv N_{\rm tw}\pmod 2,\qquad N_{\rm tw}=\sum_{g_\alpha=-1}N_\alpha ,
\label{eq:parity}
\end{equation}
which is the microscopic origin of the half-integer transverse spins of $g$-odd defect operators. Any
fermion bilinear $\psi^\dagger_\alpha\psi_\beta$ is antiperiodic around the defect if exactly one of
$\alpha,\beta$ is twisted, and periodic otherwise. The low-energy CFT operators are represented by charge-neutral fermion composites. Specifying the twist of each fermion therefore fixes the monodromy of their bilinears and of higher composites.
Table~\ref{tab:defects} summarizes the defects considered in this work. Next we describe each model in turn.

\paragraph{Ising.} The model of Ref.~\cite{Zhu:2022gjc} is written for two flavours $(\psi_\uparrow,\psi_\downarrow)$,
\begin{equation}
H=\int\!\!\int U(\mathbf x_{12})\big[n^0(\mathbf x_1)n^0(\mathbf x_2)-n^z(\mathbf x_1)n^z(\mathbf x_2)\big]-h\int n^x,
\end{equation}
where $n^{\mu}=\psi^\dagger\sigma^\mu\psi $ is density operator. We take the critical point of Ref.~\cite{Zhu:2022gjc}, $V_0=4.75$, $V_1=1$, $h=3.16$. Since the Ising symmetry exchanges $\psi_\uparrow$ and $\psi_\downarrow$ and is off-diagonal in this basis, we work in the $\mathbb Z_2$ diagonal basis $\psi_\pm=(\psi_\uparrow\pm\psi_\downarrow)/\sqrt2$. In this basis
the order parameter (Eq.~\eqref{eq:orderp}) becomes the off-diagonal  bilinear $\sigma\sim n^z=\psi_+^\dagger\psi_-+\mathrm{h.c.}$. The twist acts on $\psi_-$
only, i.e. $\psi_+$ keeps the $\Norb$ bulk orbitals while $\psi_-$ gets the $\Norb-1$ orbitals of the half-shifted grid Eq.~\eqref{eq:twisted}.  
By Eq.~\eqref{eq:parity}, $(-1)^{N_-}=(-1)^{2L_z}$: the $\mathbb Z_2$ charge is not an independent label, and each
sector is the union of all $N_-$ of the parity fixed by $2L_z$. In this context, $\mathbb Z_2$-odd defect operators
($\psi_s$) live at half-integer $L_z$ and $\mathbb Z_2$-even ones (e.g. $s$, $p^o$, $\hat D$, $t_+$) at integer $L_z$.

\paragraph{O(2).} We use the $(N+1)$-flavour model of Ref.~\cite{Guo:2025onf}: 
\begin{equation}
H=\int d^2\mathbf{x}\,\Big[n_{\rm tot}^2-U_1\sum_{a=1}^N V_a\nabla^2V_a+h_1\,\psi_0^\dagger\psi_0\Big],
\label{eq:Hguo}
\end{equation}
with a singlet flavour $\psi_0$, an O($N$)
vector $\psi_a$, and $ V_a=\psi_0^\dagger\psi_a+\psi_a^\dagger\psi_0$.
We choose $(U_1,h_1)=(0.270,-0.084)$ for O(2) and $(0.264,-0.096)$ for O(3)~\cite{Guo:2025onf}.
In the rotating basis $(\psi_+,\psi_0,\psi_-)$ (with
$\psi_\pm=(\psi_1\pm i\psi_2)/\sqrt2$),  
both the U(1) and $g=\mathrm{diag}(-1,+1,-1)$ are diagonal. 
$\psi_\pm$ are twisted, but $\psi_0$ is not. The conserved labels are the charge $Q=N_+-N_-$ and angular momentum $2L_z$. Since both
twisted flavors carry $Q=\pm1$, Eq.~\eqref{eq:parity} gives $2L_z\equiv Q\pmod2$: charged defect operators
have half-integer transverse spin, exactly as $\Psi_s$ in the bulk-to-defect expansion of $\phi_\pm\sim V_1\pm iV_2$.
Charge conjugation $C:\psi_+\leftrightarrow\psi_-$ commutes with the twist (both flavours are twisted) and labels
the $Q=0$ sectors.

\paragraph{O(3).} For O(3) we use $(\psi_+,\psi_z,\psi_-,\psi_0)$ with $\psi_\pm=(\psi_x\pm i\psi_y)/\sqrt2$ and
$S^z=(+1,0,-1,0)$ ~\cite{Guo:2025onf}. We will consider three monodromies, which are diagonal in this basis and differ only in the
twisted set (Table~\ref{tab:defects}). The Hamiltonian, the flavour tensors and the one-body terms are
identical for the three defects:
\begin{itemize}
\item \emph{Centre twist} $g=-\one$: $\psi_+,\psi_z,\psi_-$ twisted. The full O(3) is preserved, so in
addition to $(S^z,2L_z)$ we measure the total spin $S$ through the Casimir built from the O(3) generators
acting on the vector flavours. No relation between $2L_z$ and $S^z$ holds, since $\psi_z$ is twisted but neutral.
\item \emph{$R_z(\pi)$} $g=\mathrm{diag}(-1,-1,+1)$: $\psi_\pm$ twisted, $\psi_z$ untwisted. The labels are $S^z$,
$2L_z$ with $2L_z\equiv S^z\pmod 2$, and the parity $P_3=(-1)^{N_z}$ of the preserved $\mathbb Z_2$
($\phi_3\to-\phi_3$), which is independent of $L_z$ here because $\psi_z$ is untwisted.
\item \emph{Reflection} $g=\mathrm{diag}(1,1,-1)$: only $\psi_z$ twisted. Now $(-1)^{N_z}$ is the $g$-charge itself
and, as for Ising, equals $(-1)^{2L_z}$; the independent labels are $S^z$, $2L_z$ and, at $S^z=0$, the
charge conjugation $C:\psi_+\leftrightarrow\psi_-$ of the preserved O(2).
\end{itemize}

\subsection{Observables}
\label{sec:observables}
\paragraph{Spectrum.} We obtain defect scaling dimensions from the energy spectrum by \cite{Hu:2023ghk,Zhou:2023fqu}
$\Dh_n=(E^{\mathcal D}_n-E^{\mathcal D}_0)/E_R$, where $E_R=c/R$ is the CFT energy unit at the given size, $c$ the
non-universal velocity of light of the microscopic model. We never need $c$ and $R$ separately:
$E_R$ is fixed at each size from the bulk spectrum by requiring $\Delta_T=3$ (and $\Delta_J=2$ for O($N$)) in a least-squares
sense, and we refer to this choice as the bulk-velocity normalization. As an alternative we also use the displacement normalization
$\Dh_{\hat D}\equiv2$. The two normalizations differ by at most $4\%$ for the O($N$) models and by $5$--$9\%$ for
Ising at the sizes studied.

\paragraph{Bulk-to-defect OPEs.} Following Ref.~\cite{Hu:2023ghk}, with
$n_{\mathcal O}(\mathbf x)=\sum_{\alpha\beta}M^{\mathcal O}_{\alpha\beta}\psi^\dagger_\alpha\psi_\beta(\mathbf x)$ a
microscopic density with the quantum numbers of $\mathcal O$, we use
\begin{align}
a_{\mathcal O}=\frac{\langle\hat0|n_{\mathcal O}(\theta)|\hat0\rangle-\langle0|n_{\mathcal O}|0\rangle}{\langle\mathcal O|n_{\mathcal O}|0\rangle}\sin^{\Delta_{\mathcal O}}\theta,\\
b_{\mathcal O\hat{\mathcal O}}=\left|\frac{\langle\hat{\mathcal O}|n_{\mathcal O}(\theta)|\hat0\rangle}{\langle\mathcal O|n_{\mathcal O}|0\rangle}\right|\sin^{\Delta_{\mathcal O}-\Dh}\theta ,
\label{eq:ab}
\end{align}
evaluated at the equator $\theta=\pi/2$, where the distance to both poles is maximal. 
The density operators are $n^z$ ($\sigma$) and $n^x$ ($\epsilon$) for Ising \cite{Zhu:2022gjc}, $V_a$ ($\phi_a$),
$\psi_0^\dagger\psi_0$ ($S$), $\psi_+^\dagger\psi_-$ (charge-2 tensor $t$) and
$\psi_z^\dagger\psi_z-\frac{1}{3}\sum_a\psi^\dagger_a\psi_a$ ($t_{33}$) for O($N$) ~\cite{Guo:2025onf}.
A microscopic density is not a unit-normalized primary: $n_{\mathcal O}=c_{\mathcal O}\mathcal O+\sum_{\mathcal O'}c_{\mathcal O'}\mathcal O'+\dots$ contains, besides $\mathcal O$ and its descendants, all other operators with the same quantum numbers. For densities carrying a charge these corrections are expected to be captured by the finite-size extrapolation.

\paragraph{Defect-defect OPE coefficients.} In radial quantization the defect line passes through the north
pole at all times, so a bulk density at polar angle $\theta_0$ near the pole is, by the bulk-to-defect
expansion, a \emph{defect-local} operator, $n_{\mathcal O}(\theta_0)\simeq\sum_{\hat{\mathcal O}}
c_{\hat{\mathcal O}}\,\theta_0^{\Dh-\Delta}\hat{\mathcal O}+\dots$ \cite{Billo:2013jda,Gaiotto:2013nva}. Its matrix elements between defect states
are therefore defect three-point functions. With $|\hat 1\rangle,|\hat 3\rangle$ defect states and
$\hat{\mathcal O}_2$ the lowest defect operator with the quantum numbers transferred by $n_{\mathcal O}$,
\begin{equation}
C_{\hat1\hat2\hat3}=\lim_{\theta_0\to0}\;
\frac{\langle\hat1|n_{\mathcal O}(\theta_0)|\hat3\rangle}{\langle\hat2|n_{\mathcal O}(\theta_0)|\hat0\rangle}\;
(\cos\theta_0)^{-(\Dh_1-\Dh_2-\Dh_3)} .
\label{eq:C3}
\end{equation}
The cosine factor is the kinematics of the defect channel of $\langle\hat 1(\infty)\,\mathcal O(x)\,\hat 3(0)\rangle$
at $|x|=1$: the defect operator is inserted at $\tau=\cos\theta_0$ on the line and
$\langle\hat1(\infty)\hat2(\tau)\hat3(0)\rangle=C\,\tau^{\Dh_1-\Dh_2-\Dh_3}$, whereas the normalizing two-point function
carries no such factor. Corrections are of relative order $\theta_0^{\Dh_2'-\Dh_2}$ from the next defect operator in
the channel and $\theta_0^2$ from descendants. At fixed fuzzy-sphere size, the limit $\theta_0\to0$ does not resolve a continuum defect operator: twisted orbitals vanish at the poles, and for $\theta_0\lesssim\ell_B/R$ local densities are dominated by the detailed core regularization. A charged density cannot become the neutral total charge density; its leading matrix element can instead vanish or become sensitive to cutoff effects. We therefore evaluate
Eq.~\eqref{eq:C3} on the window $\theta_0\in\{\pi/12,\pi/6,\pi/4\}$, and use the corrected ratio averaged over this window as our estimate, with half its spread as a systematic error. A plateau is observed in the well-converged channels; the exception $C_{\psi\bar\psi p^o}$ is discussed in App.~\ref{app:plateau}. Only
charged densities $n_{\mathcal O}$, for which the defect operator $\hat{\mathcal O}_2$ transferred in Eq.~\eqref{eq:C3} carries a charge, are used (the order-parameter densities $n^z$, $V_a$, and $V_3$ when it is twisted or untwisted, as appropriate).
Coefficients in which $\hat{\mathcal O}_2$ is neutral follow from the permutation symmetry of the three-point coefficient. 
All coefficients are reported as absolute values.

\section{Results}
\label{sec:results}

\subsection{Ising: the $\mathbb{Z}_2$ twist defect}
\label{sec:ising}

\begin{figure}[t]
\centering
\includegraphics[width=\linewidth]{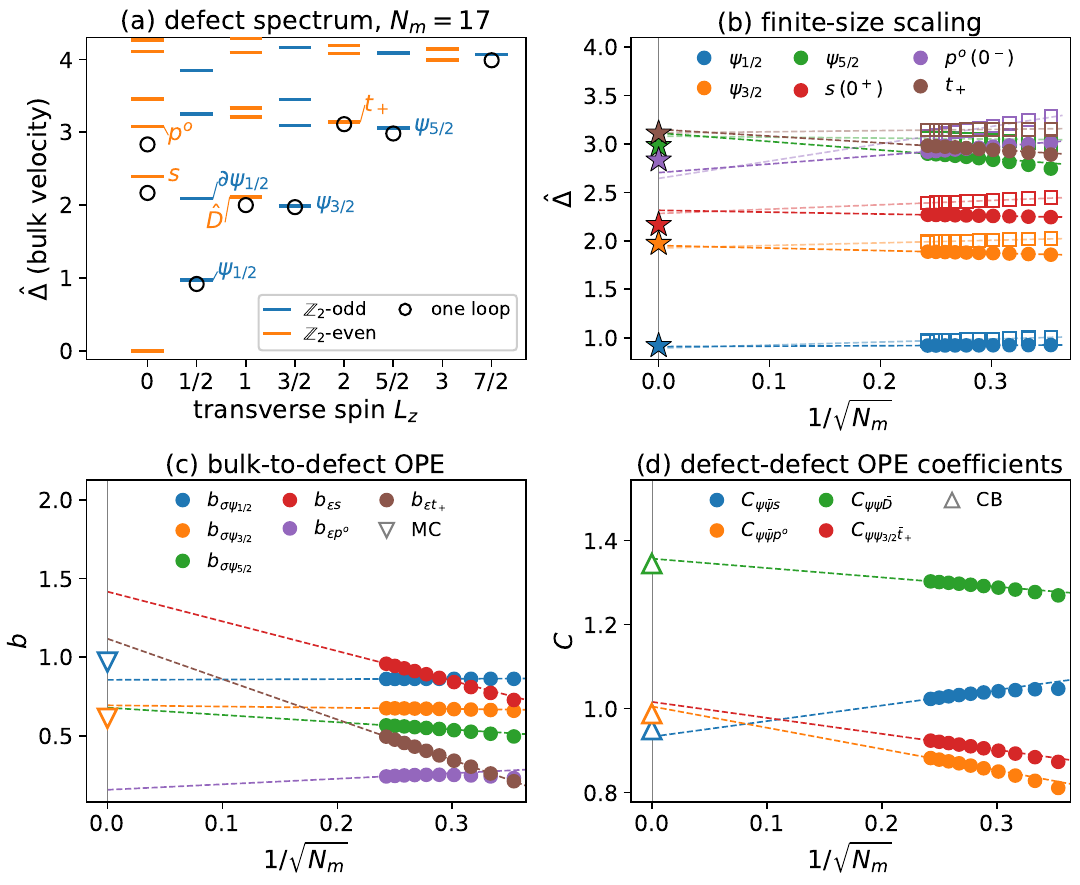}
\caption{Ising $\mathbb Z_2$ twist defect. (a) Defect spectrum versus transverse spin $L_z$ at $\Norb=17$ (
blue: $\mathbb Z_2$-odd, orange: $\mathbb Z_2$-even; open circles: one-loop perturbative values). 
(b) Finite-size scaling of the defect spectrum versus $1/\sqrt{\Norb}$ (filled circles: displacement normalization; open squares: bulk velocity normalization). Dashed lines are fits to the four largest sizes for both normalizations,  and the values from one-loop perturbation are labeled by stars. Finite-size scaling of (c) bulk-to-defect OPE and (d)
defect-defect OPE coefficients. In (c) and (d) the reference values are open markers in the colour of the corresponding data: downward triangles for Monte Carlo (MC)~\cite{Billo:2013jda} and upward
triangles for the conformal bootstrap (CB)~\cite{Gaiotto:2013nva}.
}
\label{fig:ising}
\end{figure}

\begin{table}[b]
\centering\footnotesize
\caption{Ising $\mathbb Z_2$ twist defect: extrapolated defect scaling dimensions from the fuzzy sphere (FS), compared with one loop, large transverse spin (LLMS)~\cite{Lemos:2017vnx}; Monte Carlo (MC)~\cite{Billo:2013jda}; conformal bootstrap (CB)~\cite{Gaiotto:2013nva}.}
\label{tab:ising}
\fitwidth{%
\begin{tabular}{lccccc}
\toprule
operator & FS & one loop & LLMS & MC & CB \\
\midrule
$\psi_{1/2}$ &  0.91(2) & 0.917 & 0.932 &  0.9187(6) & 0.9187 \\
$\psi_{3/2}$ & 1.94(2) & 1.972 & 1.978 &  1.99(5)& --\\
$\psi_{5/2}$ & 3.06(6) & 2.983 & 2.990 & -- & --\\
$s\,(0^+)$ &  2.31(3) & 2.167 & -- &  2.27(1) & 2.27\\
$p^{o}\,(0^-)$ & 2.76(11) & 2.833 & -- & 2.9(2) & 2.92 \\
$t_+$ & 3.11(4) & 3.111 & -- & 3.1(5) & --\\
\bottomrule
\end{tabular}}
\end{table}

\begin{table}[t]
\centering\footnotesize
\caption{Ising $\mathbb{Z}_2$ twist defect: bulk-to-defect  and defect-defect OPE coefficients. One loop, MC and bootstrap values are from Refs.~\cite{Gaiotto:2013nva,Billo:2013jda,Kos:2016ysd}.
 }
\label{tab:ising_ope}
\fitwidth{%
\begin{tabular}{lccccc}
\toprule
quantity & FS & free / GFF & one loop & MC & CB \\
\midrule
$b_{\sigma\psi_{1/2}}$ & 0.857(3) & 0.798 & 0.847 & 0.968(2) & -- \\
$b_{\sigma\psi_{3/2}}$ & 0.689(5) & 0.651 & 0.680 & 0.61(9) & -- \\
$b_{\sigma\psi_{5/2}}$ & 0.65(3) & 0.583 & -- & -- & --\\
$b_{\epsilon s}$ & 1.30(12) & -- & -- & -- & --\\
$b_{\epsilon t_+}$ & 0.95(16) & -- & -- & -- & --\\
\midrule
$C_{\psi\bar\psi s}$ & 0.96(3) & 1.000 & 0.955 & -- & 0.95\\
$C_{\psi\bar\psi p^o}$ & 0.97(6) & 0.957 & 0.987 & -- & 0.988\\
$C_{\psi\psi\bar D}$ & 1.34(4) & 1.414 & 1.382 & -- & 1.345\\
$C_{\psi_{1/2}\psi_{3/2}\bar t_+}$ & 0.99(3) & 1.000 & -- & -- & --\\
$C_{\bar\psi_{3/2}\psi_{1/2}\hat D}$ & 0.31(7) & -- & -- & -- & --\\
\bottomrule
\end{tabular}}
\end{table}

\paragraph{Spectrum.} 
Fig.~\ref{fig:ising}(a) presents representative operator spectrum of $\mathbb{Z}_2$ twist defect in the Ising CFT. The most intriguing feature is observation of defect operators $\psi_{s}$ with half-integer transverse spins $s$ up to $7/2$. The emergence of half-integer spins is a direct consequence of operators being odd under the monodromy (see Eq. \eqref{eq:bdexp}).  
We resolve defect primaries and their descendants, providing evidence for the conformal nature of this monodromy defect. 
Table~\ref{tab:ising} summarizes the six twist defect operators of 3D Ising CFT. The lowest
defect operator extrapolates to $\Dh_{\psi_{1/2}}=0.91(2)$, consistent with the Monte Carlo estimate
$0.9187(6)$~\cite{Billo:2013jda} and the bootstrap $0.9187$~\cite{Gaiotto:2013nva}. 
We also identify the defect singlet $s$, spin-3/2 defect operator $\psi_{3/2}$ and the spin-2 operator
$t_+$. 
The pseudoscalar $p^o$, $2.76(11)$, lies slightly below the  bootstrap values and closer to one loop ($2.833$), but is compatible with Monte Carlo within the combined uncertainty. 
Higher transverse spins converge from below, because $\psi_s$ needs orbitals with $|m|\simeq s$, close to the edge
of the available range.

\paragraph{Bulk-to-defect OPEs.}
The coupling of $\sigma$ to the lowest mode,
$b_{\sigma\psi_{1/2}}=0.857(3)$ (Table~\ref{tab:ising_ope}), has a weak size dependence at the equator; the two adopted extrapolation forms give $0.856$ and $0.859$ (App.~\ref{app:fits}). 
It lies $7\%$ above the free value
$\sqrt{2/\pi}=0.798$ and is close to the one-loop value $0.847$~\cite{Gaiotto:2013nva}, but it is $11\%$ below the Monte Carlo value $0.968(2)$~\cite{Billo:2013jda}. 
For the next modes we find $b_{\sigma\psi_{3/2}}=0.689(5)$, also close to the one loop result $0.680$.
We also identify $b_{\sigma\psi_{5/2}}= 0.65(3)$, and the couplings of $\epsilon$ to the $\mathbb Z_2$-even operators $s$ and $t_+$, $b_{\epsilon s}=1.30(12)$ and $b_{\epsilon t_+}=0.95(16)$, which have not been obtained by other methods. Since the microscopic density $n^x$ also contains the energy density $T_{\tau\tau}$, $b_{\epsilon s},b_{\epsilon t_+}$ should be read as indicative.

\paragraph{Defect-defect OPE coefficients.} 
The defect-defect coefficients in Fig.~\ref{fig:ising}(d) converge smoothly with $\Norb$ and the $\theta_0$ dependence of the ratio is illustrated in App.~\ref{app:plateau}. The displacement
coefficient in $\psi_{1/2}\times\psi_{1/2}$, $C_{\psi\psi\bar D}=1.34(4)$, agrees with the bootstrap value
$1.345$~\cite{Gaiotto:2013nva} and lies below the generalized free field value $\sqrt2$ and the one-loop value $1.382$.
$C_{\psi\bar\psi s}=0.96(3)$ and $C_{\psi\bar\psi p^o}=0.97(6)$ are compatible with the bootstrap values $0.95$ and $0.988$ (see Table~\ref{tab:ising_ope}). 
Two further coefficients are
predictions: $C_{\psi_{1/2}\psi_{3/2}\bar t_+}=0.99(3)$ and $C_{\bar\psi_{3/2}\psi_{1/2}\hat D}=0.31(7)$.

Based on the above data, we deduce that the twist defect of the Ising CFT is close to a generalized free field of modes $\psi_s$ dressed by a weak repulsive contact interaction. The evidence includes: (i) In the single-mode tower $\Dh_s-s-\Delta_\sigma$ is $-0.11$ at $s=1/2$ and $-0.08$ at
$s=3/2$ with the approach to zero at large $s$ expected from the large-spin expansion~\cite{Lemos:2017vnx}. (ii) Two-mode operators are pushed up from $\Dh_{s_1}+\Dh_{s_2}$: the singlet
$s=[\psi_{1/2}\bar\psi_{1/2}]$ by $0.49(4)$, the displacement by $2-2\Dh_{1/2}=0.18(4)$ (fixed by $\Dh_{\hat D}=2$) and $t_+=[\psi_{1/2}\psi_{3/2}]$ by
$0.26(5)$. The one-loop contact term (see Appendix Sec. \ref{sec:oneloop}), gives $0.33$, $0.17$ and $0.22$: it reproduces the pattern and the
displacement exactly, and underestimates the singlet by about $30\%$. (iii) The OPE coefficients deviate from generalized free field by at most
$5\%$, with the largest depletion in the identical-mode channel $\psi_{1/2}\psi_{1/2}\to\bar D$. The leading $\epsilon$-expansion thus captures much of the single-mode and OPE structure, although some differences exceed the quoted fit errors; it also underestimates the
two-mode singlet energy, much as it underestimates $\Delta_\epsilon$ in the bulk.


\subsection{O(2): the $\pi$-monodromy defect}
\label{sec:o2}
\begin{figure}[t]
\centering
\includegraphics[width=\linewidth]{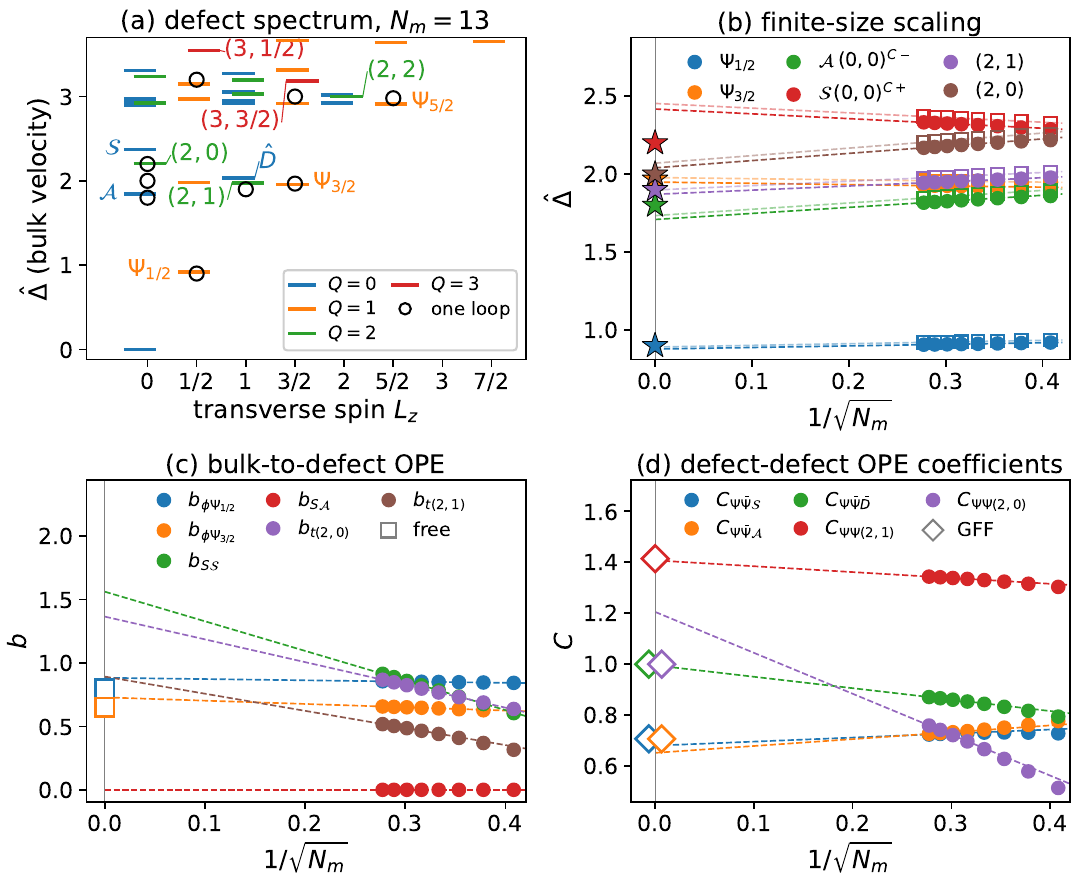}
\caption{O(2) $\pi$-monodromy defect. (a) Defect spectrum where each defect operator carries U(1) charge $Q$ and transverse spin $L_z\in\mathbb Z/2$. 
Open circles are results from one-loop. 
Finite-size scaling of  (b) defect spectrum, (c) bulk-to-defect OPEs and (d) defect-defect OPE coefficients. Reference values are from one-loop calculation (stars in (b)), free theory (squares in (c)) and GFF (diamonds in (d)). 
}
\label{fig:o2}
\end{figure}
\begin{table}[t]
\centering\footnotesize
\caption{O(2) $\pi$-monodromy defect. Labels $(Q,L_z)^{C}$ denote U(1) charge $Q$, transverse spin $L_z$ and charge conjugation $C$.
The second (third) column is the finite-size data obtained by bulk normalization (displacement operator normalization). The fourth column is the extrapolated data on fuzzy sphere (FS), and fifth column is from one-loop calculations (Appendix Sec.\ref{sec:oneloop}). One-loop values of three-mode operators are the pairwise sums of the two-mode contact terms. (n.c.): not converged, extrapolation indicative only. }
\label{tab:o2}
\fitwidth{%
\begin{tabular}{lcccc}
\toprule
operator & $N_m=13$ (bulk) & $N_m=13$ ($\hat D$) & FS & one loop \\
\midrule
$\Psi_{1/2}$ & 0.922 & 0.907 & 0.89(1) & 0.900 \\
$\Psi_{3/2}$ & 1.958 & 1.926 & 1.96(2) & 1.967 \\
$\Psi_{5/2}$ & 2.911 & 2.865 & 3.24(15)  & 2.980 \\
$\Psi_{7/2}$ & 3.654 & 3.596 & 4.82(47) (n.c.)  & 3.986 \\
$\mathcal{A}\,(0,0)^{C-}$ & 1.847 & 1.817 & 1.75(4) & 1.800 \\
$\mathcal{S}\,(0,0)^{C+}$ & 2.370 & 2.332 & 2.41(4) & 2.200 \\
$\hat D$ & 2.032 & 2 (input) & 2 (exact) & 2 \\
$(2,1)$ & 1.976 & 1.945 & 1.91(3) & 1.900 \\
$(2,0)$ & 2.204 & 2.169 & 2.09(5) & 2.000 \\
$(2,2)$ & 3.002 & 2.954 & 3.10(7) & 3.000 \\
$(3,\frac{3}{2})$ & 3.183 & 3.132 & 3.05(6) & 3.000 \\
$(3,\frac{1}{2})$ & 3.546 & 3.489 & 3.36(8) & 3.200 \\
\bottomrule
\end{tabular}}
\end{table}
\begin{table}[t]
\centering\footnotesize
\caption{O(2) $\pi$-monodromy defect: bulk-to-defect and defect-defect OPE coefficients.}
\label{tab:o2_ope}
\fitwidth{%
\begin{tabular}{lccc}
\toprule
quantity & $N_m=13$ & FS & free / GFF \\
\midrule
$b_{\phi\Psi_{1/2}}$ & 0.8561 & 0.876(7) & 0.798  \\
$b_{\phi\Psi_{3/2}}$ & 0.6571 & 0.71(2) & 0.651  \\
$b_{t(2,0)}$ & 0.8681 & 1.23(13)  & --  \\
$b_{t(2,1)}$ & 0.5181 & 0.79(10)  & --  \\
$b_{S\mathcal S}$ & 0.9149 & 1.39(17) (n.c.)  & -- \\
$b_{S\mathcal A}$ ($C$-forbidden) &  0.0000 & 0.0000 & 0 (exact) \\
\midrule
$C_{\Psi\bar\Psi\mathcal S}$ & 0.724 & 0.69(2) & 0.707  \\
$C_{\Psi\bar\Psi\mathcal A}$ & 0.727 & 0.67(2) & 0.707  \\
$C_{\Psi\bar\Psi\bar D}$ & 0.871 & 0.96(5) & 1.000  \\
$C_{\Psi\Psi(2,1)}$ & 1.344 & 1.39(3) & 1.414  \\
$C_{\Psi_{1/2}\Psi_{-1/2}(2,0)}$ & 0.759 & 1.09(14)  & 1.000  \\
$C_{\Psi_{1/2}\Psi_{3/2}(2,2)}$ & 0.193 & 0.24(2) (n.c.)  & --  \\
\bottomrule
\end{tabular}}
\end{table}

The results for the $\pi$-rotation of the O(2) CFT are shown in Fig. \ref{fig:o2},   
where a series of defect operators can be identified. 
The single-mode tower $\Psi_s$ are estimated to be $\Dh_{\Psi_{1/2}}=0.89(1)$ and
$\Dh_{\Psi_{3/2}}=1.96(2)$ (see Table~\ref{tab:o2}). The lowest neutral $L_z=0$ operator is
\emph{charge-conjugation odd}, $\mathcal A=\epsilon_{ab}\psi^a_{1/2}\psi^b_{-1/2}$, at $1.75(4)$. The charge-2 operators $(Q,L_z)=(2,1)$ and $(2,0)$ are at
$1.91(3)$ and $2.09(5)$, the charge-3 operators at $3.05(6)$ and $3.36(8)$.
The one-loop calculation captures the ordering of these levels, with quantitative deviations in several composite channels. The spectra provide non-perturbative estimates for these operator dimensions.

Tab.~\ref{tab:o2_ope} summarizes 6 bulk-to-defect OPEs and 6 defect-to-defect OPEs.  
In comparison with those from the Ising CFT, $b_{\phi\Psi_{1/2}}=0.876(7)$ and $b_{\phi\Psi_{3/2}}=0.71(2)$  are within a few percent of the Ising values. And many others are close to those from the free theory. For instance, both $C_{\Psi\bar\Psi\mathcal S}=0.69(2)$ and $C_{\Psi\bar\Psi\mathcal A}=0.67(2)$ are close to the GFF value $1/\sqrt2=0.707$, and the identical-mode channel $\Psi_{1/2}\Psi_{1/2}\to(2,1)$ gives $1.39(3)$, compatible with $\sqrt2$ from the GFF.

No symmetry-preserving relevant defect operator is resolved in the low-energy spectrum: every resolved neutral,
$C$-even scalar has $\Dh>1$. The only resolved relevant multiplet, $\Psi_{1/2}$, carries charge and transverse spin, which may yield the ``monodromy pinning'' deformation~\cite{Kravchuk:2025mpd}. To quantify how two defect modes interact we use the interaction energy $\Dh_{[12]}-\Dh_1-\Dh_2$, the shift of a two-mode operator from the sum of its constituents, which vanishes for a generalized free field. At one loop it is the contact term Eq.~\eqref{eq:oneloop2}: the quartic coupling $(\phi\cdot\phi)^2$ is symmetric in its flavour indices, so it repels flavour-symmetric pairs, with a strength set by the flavour weight $T_R$, and does not act on antisymmetric pairs. The O(2) data follow this pattern (see Fig.~\ref{fig:physics}): the antisymmetric operator $\mathcal A$ has an interaction energy consistent with zero, $-0.03(5)$ ($T_{\mathcal A}=0$), while the symmetric channels are repelled, $(2,1)$ by $0.12(4)$ [one loop $0.10$], $(2,0)$ by $0.31(6)$ [$0.20$] and $\mathcal S$ by $0.63(5)$ [$0.40$], in the order of their flavour weights but $1.2$--$1.6$ times more strongly than at one loop. 
The charge-2 operator
$(2,0)$ and the displacement are degenerate at one loop; non-perturbatively the displacement stays at $2$ while $(2,0)$ moves to
$2.09(5)$, so the degeneracy is an accident of the leading order.

\subsection{O(3): three inequivalent monodromy defects}
\label{sec:o3}

\begin{figure*}[t]
\centering
\includegraphics[width=\textwidth]{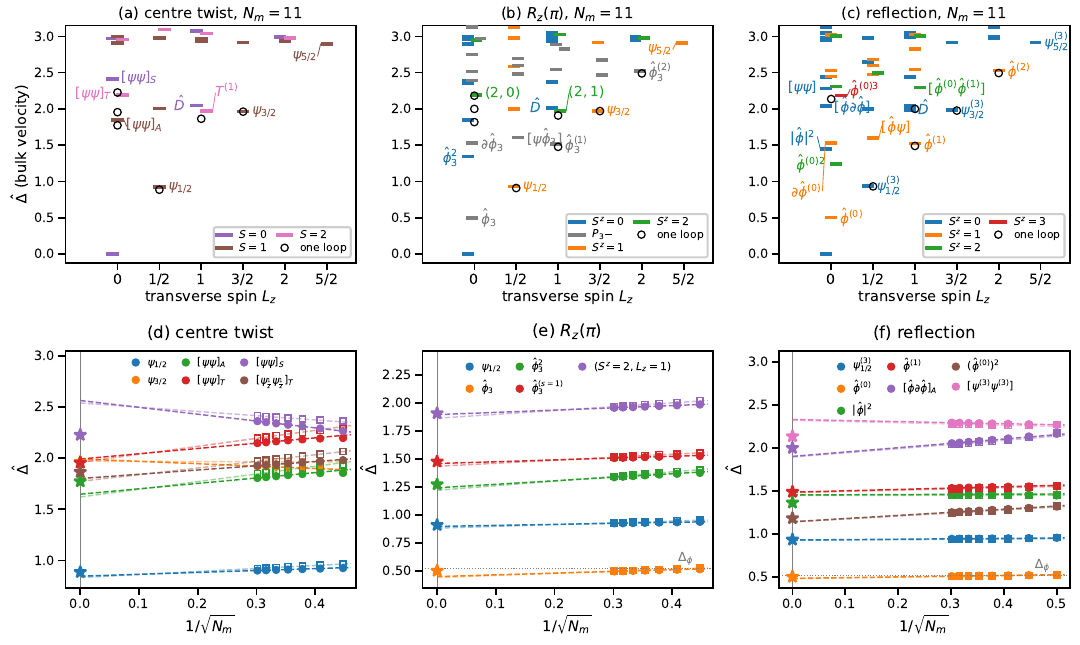}
\caption{Defect spectra of the three O(3) monodromy defects. (Top): Spectra versus transverse spin $L_z\in\mathbb Z/2$ at $\Norb=11$ for
(a) the centre twist $g=-\one$ (colour: total spin $S$), (b) $R_z(\pi)$ (colour: $S^z$; grey: $P_3$-odd) and (c) the reflection
$\phi_3\to-\phi_3$ (colour: $S^z$). 
$T^{(1)}$ is the identical-mode quintet $[\psi_{1/2}\psi_{1/2}]_T$ at $L_z=1$, $(2,L_z)$ the lowest $S^z=2$ operator of
transverse spin $L_z$, $\psi^{(3)}_s$ the twisted modes of $\phi_3$ and $\hat\phi^{(s)}$ the modes of the untwisted doublet.
Open circles show the perturbative references; for the partial twists these use the free zero-mode prescription specified in Tables~\ref{tab:o3r} and~\ref{tab:o3z}.
(Bottom): finite-size scaling of the corresponding spectra for (d) the centre twist, (e) $R_z(\pi)$ and (f) the reflection. The dotted line in (e) and (f) is the bulk value of $\Delta_\phi$, below which the untwisted $s=0$ modes lie.
}
\label{fig:o3spec}
\end{figure*}

O(3) admits three inequivalent $\mathbb Z_2$ monodromies, distinguished by the number $n_{\rm tw}=3,2,1$ of twisted
components (Sec.~\ref{sec:bases}). They share the same microscopic Hamiltonian and differ only in the twisted
flavours (see Tab. \ref{tab:defects}): The centre twist has a purely twisted spectrum; the $R_z(\pi)$ defect has one untwisted component, a
$\mathbb Z_2$-odd singlet; the reflection has two untwisted components forming a charged O(2) doublet. The latter two
contain an untwisted $s=0$ mode for which the attractive reference-background potential gives an imaginary radial exponent. This signals the failure of a fixed free zero-mode boundary condition, rather than a complex scaling dimension of the unitary numerical model. App.~\ref{sec:untwtheory} discusses this mass flow, and its relation to the no-explicit-pinning prescription used in all our calculations.
Their defect spectra and the finite-size scaling of the spectra are shown in Fig.~\ref{fig:o3spec}; the bulk-to-defect couplings,
one-point functions and defect-defect OPE coefficients of all three defects are collected in App.~\ref{app:O3}.

\subsubsection{Centre twist $g=-\one$}
\label{sec:o3c}

\begin{table}[t]
\centering\footnotesize
\caption{O(3) centre twist. $A$, $T$, $S$: antisymmetric, symmetric-traceless and singlet two-mode operators.}
\label{tab:o3c}
\fitwidth{%
\begin{tabular}{lcccc}
\toprule
operator & $N_m=11$ (bulk) & $N_m=11$ ($\hat D$) & FS & one loop \\
\midrule
$\psi_{1/2}$ & 0.922 & 0.902 & 0.86(2) & 0.886 \\
$\psi_{3/2}$ & 1.964 & 1.922 & 1.97(2) & 1.962 \\
$\psi_{5/2}$ & 2.894 & 2.831 & 3.36(20) & 2.977 \\
$\hat D$ & 2.045 & 2 (input) & 2 (exact) & 2 \\
$[\psi\psi]_A$ & 1.847 & 1.806 & 1.68(6) & 1.773 \\
$[\psi\psi]_T$ & 2.195 & 2.147 & 2.02(6) & 1.955 \\
$[\psi\psi]_S$ & 2.414 & 2.362 & 2.51(5) & 2.227 \\
$[\psi_{\frac{1}{2}}\psi_{\frac{1}{2}}]_T$ & 1.969 & 1.926 & 1.83(5) & 1.864 \\
\bottomrule
\end{tabular}}
\end{table}

The centre twist $g=-\one$ twists all three components, preserves the full O(3), and is the O(3) analogue of
the Ising twist and of the O(2) $\pi$-monodromy; it is the defect with the largest number of twisted components, $n_{\rm tw}=M=3$, and
hence with the strongest one-loop attraction, $\mu_{\rm tw}\propto(n_{\rm tw}+2)/(M+8)=5/11$ (compared with $4/10$ for O(2) and $3/9$ for
Ising). $g$-odd defect operators carry half-integer and $g$-even ones integer transverse spin; all organize into O(3) multiplets. The single-mode tower is a triplet $\psi^a_s$ with
$\Dh_{\psi_{1/2}}=0.86(2)$ and $\Dh_{\psi_{3/2}}=1.97(2)$ (see Table~\ref{tab:o3c}). The product of two triplets decomposes into $1\oplus3\oplus5$, and the two-mode operators
at $L_z=0$, $[\psi_{1/2}\psi_{-1/2}]$, appear as a singlet $S$, an antisymmetric triplet $A$ and a symmetric traceless quintet $T$, in
the order $A<T<S$ predicted by the flavour weights of one-loop calculation (see Appendix Sec. \ref{sec:oneloop}): $[\psi\psi]_A=1.68(6)$,
$[\psi\psi]_T=2.02(6)$, $[\psi\psi]_S=2.51(5)$. The symmetric-traceless component of $[\psi_{1/2}\psi_{1/2}]$ at $L_z=1$ is a quintet at $1.83(5)$; the singlet is the displacement, while the antisymmetric component vanishes.

The centre twist is the cleanest illustration of the flavour structure of the contact interaction, because
all three O(3) channels of the same pair of modes are resolved at the same $L_z$. Their interaction energies,
$\Dh_{[\psi\psi]}-2\Dh_{\psi_{1/2}}$, are $-0.03(7)$ for $A$, $0.31(7)$ for $T$ and $0.79(6)$ for $S$: the antisymmetric channel is
consistent with no interaction, as required by the symmetric flavour tensor of the quartic coupling ($T_A=0$), and the ratio
$T:S=0.39(10)$ agrees with the one-loop ratio of flavour weights $\frac{2}{3}:\frac{5}{3}=0.40$, while the absolute values are $1.7$ and $1.8$ times
the one-loop estimation $0.18$ and $0.46$. The identical-mode quintet at $L_z=1$ has interaction energy $0.11(6)$ [one loop $0.09$], the
displacement exactly $2-2\Dh_{\psi_{1/2}}=0.28$ [one loop $0.23$]. Across the three $\mathbb Z_2$ defects of the Ising, O(2) and O(3) CFTs the
twisted sector is thus remarkably uniform: the lowest mode lies below $\frac{1}{2}+\Delta_\phi$ by an amount that grows with
$(n_{\rm tw}+2)/(M+8)$ ($0.11$, $0.13$, $0.16$), the bulk-to-defect coupling grows in the same order ($0.86$, $0.88$, $0.88$), symmetric
two-mode channels are repelled by $1.2$--$1.8$ times the one-loop contact term, antisymmetric ones almost vanish, and the
identical-mode OPE coefficients are $1$--$5\%$ below $\sqrt2$. The resolved spectrum has a single relevant multiplet, the triplet $\psi_{1/2}$
($\Dh=0.86$), which carries transverse spin $\frac{1}{2}$ and O(3) charge; its insertion breaks O(3) to O(2) and $SO(2)_\perp$ to a diagonal
subgroup and is the O(3) version of monodromy pinning~\cite{Kravchuk:2025mpd}. The lowest O(3)-invariant defect operator resolved
is $[\psi\psi]_S$ at $2.51(5)$, far above marginality, and the lowest $S^z=0$ level in a nontrivial O(3) multiplet, $[\psi\psi]_A$, lies at $1.68(6)$:
within the sizes studied no symmetry-preserving relevant operator appears, and the centre twist is the most robustly stable
of the five $\mathbb Z_2$ defects. Unlike the partial twists, the centre twist has no untwisted zero mode and hence no associated classically marginal mass coupling. This removes that particular weak-coupling instability, without implying independence from arbitrary strong core perturbations. These numbers -- in particular the dimensions and OPE coefficients (App.~\ref{app:O3}) of the three $[\psi\psi]$ channels --
are the first non-perturbative data for a monodromy defect of the O(3) CFT and are directly comparable to a future defect
bootstrap along the lines of Ref.~\cite{Gaiotto:2013nva}.

\subsubsection{$R_z(\pi)$: one untwisted component}
\label{sec:o3r}

\begin{table}[t]
\centering\footnotesize
\caption{O(3) $R_z(\pi)$ defect. $\hat\phi_3^{(s)}$: modes of the untwisted component. The last column gives reference-background estimates from App.~\ref{sec:oneloop}, with the untwisted $s=0$ mode assigned its free value $\frac{1}{2}$ and the stated contact terms included for composites. These are not controlled fixed-point predictions for the partial twist; in particular, values involving $s=0$ do not include its RG flow (App.~\ref{sec:untwtheory}).}
\label{tab:o3r}
\fitwidth{%
\begin{tabular}{lcccc}
\toprule
operator & $N_m=11$ (bulk) & $N_m=11$ ($\hat D$) & FS & one loop \\
\midrule
$\psi_{1/2}$ & 0.929 & 0.924 & 0.90(1) & 0.909 \\
$\psi_{3/2}$ & 1.973 & 1.963 & 2.03(3) & 1.970 \\
$\psi_{5/2}$ & 2.905 & 2.890 & 3.43(21) (n.c.)  & 2.982 \\
$\hat\phi_3$ & 0.497 & 0.494 & 0.46(2) & 0.500 \\
$\hat\phi_3^{(s=1)}$ & 1.516 & 1.508 & 1.46(3) & 1.477 \\
$\hat\phi_3^{(s=2)}$ & 2.520 & 2.507 & 2.64(6)  & 2.489 \\
$\hat\phi_3^2$ & 1.344 & 1.337 & 1.26(4) & 1.273 \\
$\hat D$ & 2.011 & 2 (input) & 2 (exact) & 2 \\
$(S^z{=}2,L_z{=}1)$ & 1.968 & 1.958 & 1.90(3) & 1.909 \\
$(S^z{=}2,L_z{=}0)$ & 2.192 & 2.181 & 2.08(5) & 2.000 \\
$\psi_{1/2}\hat\phi_3$ & 1.603 & 1.595 & 1.48(5) & 1.530 \\
\bottomrule
\end{tabular}}
\end{table}

The $R_z(\pi)$ defect twists $(\phi_1,\phi_2)$ and preserves O(2)$\times\mathbb Z_2$. Its twisted
sector is numerically consistent with the O(2) $\pi$-monodromy, as expected, since at one loop the twisted sector depends only on $(n_{\rm tw}+2)/(M+8)$, which is nearly the same for the two defects: $\Dh_{\psi_{1/2}}=0.90(1)$, $\Dh_{\psi_{3/2}}=2.03(3)$, $(S^z,L_z)=(2,1)$ at $1.90(3)$ and $(2,0)$ at
$2.08(5)$ (Table~\ref{tab:o3r}).
The untwisted component $\phi_3$ gives integer-spin modes $\hat\phi_3^{(s)}$.
The dimension of $s=0$ mode decreases slowly, and extrapolates to $0.46(2)$, below the free value $1/2$. The leading-order expansion about the free background does not determine this dimension. The $s=1$ value is close to the perturbative reference, while $s=2$ has a larger extrapolation dependence. The enhanced zero-mode correlations are qualitatively consistent with the negative mass generated near four dimensions, but their three-dimensional exponent is determined here from the spectrum rather than from the weak-coupling RG. The closest Monte Carlo comparison in Ref.~\cite{MCheng2026a} is its zero-explicit-pinning result: $0.409(11)$ for the untwisted mode and $0.920(36)$ for the twisted mode, compared with our $0.46(2)$ and $0.90(1)$. The twisted estimates agree within their uncertainties, whereas the untwisted estimates differ by about two combined quoted uncertainties. The stronger-pinning value $0.194(6)$  at $\tilde c=-1.8$ characterizes another microscopic choice. Universality would require agreement if both choices reach the same fixed point; that premise and the remaining finite-size effects must be tested.

The untwisted    mode lies near the free zero-mode reference  $\frac{d-2}{2}=\frac{1}{2}$,   and its bulk-to-defect coupling remains close to the trivial-defect value $1$ (App.~\ref{app:O3}); its composites nevertheless have substantial interaction energies and depleted OPE coefficients.    The value $\frac{1}{2}$ is not a unitarity bound for a one-dimensional defect scalar. No  symmetry-preserving relevant    scalar is resolved:  $\hat\phi_3$ is odd under the preserved $\mathbb Z_2$   , while the even level denoted $\hat\phi_3^2$ has dimension $1.26(4)>1$. Its interaction energy is approximately  $0.34$   , compared with  $2\hat\Delta_{\hat\phi_3}\simeq0.92$   , and alternative  three-parameter fits  still  give $1.18$--$1.20$   . This supports stability in  the    resolved singlet channel. At a defect fixed point, the dimension of a mass eigenoperator is set by an eigenvalue of the RG stability matrix and need not equal twice the dimension of $\hat\phi_3$. The bare composite can mix with other singlets, so its measured dimension is not simply related to the perturbative mass flow of Ref.~\cite{MCheng2026a}. The weak-coupling runaway found there near $d=4$ leaves the three-dimensional endpoint undetermined and is compatible with either a strong-coupling conformal regime or, in principle, a line with spontaneously broken $\mathbb Z_2$. Our spectra support the conformal interpretation and show no resolved symmetry-breaking signal at the available sizes; a possible crossover requires larger-size tests. The  relevant symmetry-breaking deformations $\psi_{1/2}$ ($\simeq0.9$) and $\hat\phi_3$ ($\simeq0.5$)    instead provide starting points for linear pinning flows~\cite{Yang:2026onl}, distinct from  the symmetry-preserving quadratic mass discussed above.

\subsubsection{Reflection $\phi_3\to-\phi_3$: an untwisted charged doublet}
\label{sec:o3z}

\begin{table}[t]
\centering\footnotesize
\caption{O(3) reflection defect $g=\mathrm{diag}(1,1,-1)$. $\psi^{(3)}_s$: twisted modes of $\phi_3$; $\hat\phi^{(s)}$: modes of the untwisted O(2) doublet.    The last column uses the same reference-background prescription and qualifications  as Table~\ref{tab:o3r}, with the untwisted $s=0$ mode    assigned  its free value $\frac{1}{2}$. 
}
\label{tab:o3z}
\fitwidth{%
\begin{tabular}{lcccc}
\toprule
operator & $N_m=11$ (bulk) & $N_m=11$ ($\hat D$) & FS & one loop \\
\midrule
$\psi^{(3)}_{1/2}$ & 0.936 & 0.942 & 0.930(6) & 0.932 \\
$\psi^{(3)}_{3/2}$ & 1.982 & 1.995 & 2.05(3) & 1.977 \\
$\psi^{(3)}_{5/2}$ & 2.916 & 2.936 & 3.47(21) (n.c.)  & 2.986 \\
$\hat\phi^{(0)}$ & 0.504 & 0.507 & 0.487(8) & 0.500 \\
$\hat\phi^{(1)}$ & 1.523 & 1.534 & 1.50(2) & 1.489 \\
$\hat\phi^{(2)}$ & 2.530 & 2.547 & 2.67(6)  & 2.494 \\
$|\hat\phi|^2$ & 1.450 & 1.460 & 1.452(5) & 1.364 \\
$[\hat\phi\partial\hat\phi]_A$ & 2.042 & 2.055 & 1.94(4) & 2.000 \\
$[\psi^{(3)}\psi^{(3)}]$ & 2.280 & 2.295 & 2.32(2) & 2.136 \\
$\hat D$ & 1.987 & 2 (input) & 2 (exact) & 2 \\
$(\hat\phi^{(0)})^2$ & 1.243 & 1.251 & 1.17(3) & 1.182 \\
$[\hat\phi^{(0)}\hat\phi^{(1)}]_{S^z=2}$ & 2.294 & 2.309 & 2.22(4) & 2.171 \\
$[\hat\phi^{(0)}\psi^{(3)}_{1/2}]$ & 1.602 & 1.612 & 1.53(3) & 1.553 \\
$(\hat\phi^{(0)})^3$ & 2.184 & 2.198 & 2.01(7)  & 2.045 \\
\bottomrule
\end{tabular}}
\end{table}

The reflection $g=\mathrm{diag}(1,1,-1)$ twists only $\phi_3$ ($n_{\rm tw}=1$) and preserves
O(2)$\times\mathbb Z_2$, with the O(2) acting on the untwisted doublet. In Tab.~\ref{tab:o3z},
we find the twisted modes $\psi^{(3)}_s$ carry no U(1) charge, and 
the untwisted modes $\hat\phi^{(s)}$ carry U(1) charge $\pm1$ and integer spin.
Additionally, the lowest resolved non-identity defect primary
with $L_z=0$ that is invariant under the full preserved
O(2)$\times\mathbb Z_2$ symmetry is
$|\hat\phi|^2=\hat\phi^{(0)}\hat\phi^{(0)\dagger}$
at $1.452(5)$. Its interaction energy, $1.452-2\times0.487=0.48(1)$, is $1.3$ times the
one-loop contact term $0.36$ (with the identical-mode factor $\frac{1}{2}$ appropriate for the two components of the same $s=0$
mode), and the charge-2 operator $(\hat\phi^{(0)})^2$ lies at $1.17(3)$, with an interaction energy of $0.20(3)$ vs.\ $0.18$. The ratio of
the two, $2.4$, is close to the one-loop flavour-weight ratio $T_S/T_T=2$. The lowest $C$-odd neutral operator is the first
derivative double-trace $[\hat\phi\partial\hat\phi]_A=\hat\phi^{(0)}\partial_\tau\hat\phi^{(0)\dagger}-(\partial_\tau\hat\phi^{(0)})\hat\phi^{(0)\dagger}$, the analogue of the Ising
$p^o$; it sits at $1.94(4)$, consistent with $2\Dh_{\hat\phi^{(0)}}+1$ (interaction energy $-0.03(5)$), as expected for an antisymmetric
channel. The mixed operator $[\hat\phi^{(0)}\psi^{(3)}_{1/2}]$ is at $1.53(3)$, with interaction energy $0.12(3)$ vs.\ the one-loop $0.12$
(flavour weight $T=\frac{1}{3}$).    Several low-lying untwisted and mixed levels are close to these  reference-background    estimates, while higher-spin levels show larger deviations. This comparison illustrates the contact-interaction pattern, without determining the zero-mode fixed-point boundary condition .

In the resolved low-energy spectrum the reflection defect has no symmetry-preserving relevant operator:
the lowest resolved non-identity scalar invariant under the full
preserved O(2)$\times\mathbb Z_2$ symmetry,
$|\hat\phi|^2$, has $\Dh=1.45>1$. Its two relevant deformations are both symmetry breaking: the
$g$-odd $\psi^{(3)}_{1/2}$ ($0.93$) and the charged doublet $\hat\phi^{(0)}$ ($\simeq0.5$). Adding the latter pins the direction of
$(\phi_1,\phi_2)$ on the line and breaks O(2), a monodromy generalization of the magnetic line defect. Together with $R_z(\pi)$, the
reflection shows a consistent picture of the untwisted $s=0$ channel: the single mode stays close to the    free zero-mode reference  $\frac{1}{2}$, with a
bulk-to-defect coupling close to the trivial-defect value, while its composites are among the most strongly interacting operators
identified in this work, both in their dimensions and in their OPE coefficients. For    this defect the zero-explicit-pinning Monte Carlo estimates ~\cite{MCheng2026a}, $0.472(6)$    for the untwisted and  $0.934(36)$  for the twisted mode, are close to our $0.487(8)$ and $0.930(6)$, respectively. Their values at $\tilde c=-1.75$, $0.411(6)$  and $0.976(66)$   , again probe a different core coupling. The reflection is a useful setting for such a comparison because the untwisted sector has continuous O( 2   ) symmetry rather than the discrete $\mathbb Z_2$ of $R_z(\pi)$ .

\subsection{Common features to all defects}
\label{sec:lessons}

\begin{figure}[t]
\centering
\includegraphics[width=\linewidth]{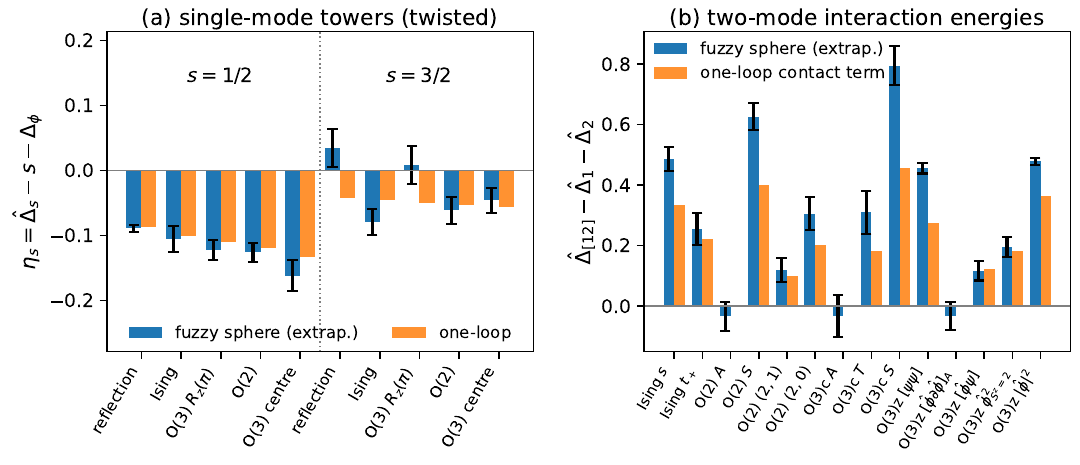}
\caption{(Left) (a) Anomalous part $\eta_s=\hat\Delta_s-s-\Delta_\phi$ of the twisted single-mode towers for the five $\mathbb Z_2$ monodromy defects, at transverse spin $s=1/2$ and $s=3/2$ (blue: fuzzy-sphere extrapolations with their uncertainties; orange: one loop, Eq.~\eqref{eq:eqmu}). The defects are ordered by the one-loop attraction $\mu_{\rm tw}\propto(n_{\rm tw}+2)/(M+8)$, which the data follow at $s=1/2$. At $s=3/2$ the anomalous parts shrink as $1/|s|$, and the one-loop spread across the five defects ($0.015$) is below the extrapolation uncertainty of $\psi_{3/2}$ ($0.02$–$0.03$); the data are consistent with one loop but cannot resolve the ordering there. (b) Interaction energies of two-mode defect operators, $\hat\Delta_{[12]}-\hat\Delta_1-\hat\Delta_2$ (for the derivative operator $[\hat\phi\partial\hat\phi]_A$: $\hat\Delta-2\hat\Delta_{\hat\phi^{(0)}}-1$), compared with the one-loop contact term, Eq.~\eqref{eq:oneloop2}.
}
\label{fig:physics}
\end{figure}

Taken together, the    spectra and OPE coefficients exhibit a common pattern: a generalized-free-field reference supplemented  by a flavour-symmetric contact interaction    captures much of the twisted-mode and composite structure. The untwisted zero modes require the additional discussion in App.~\ref{sec:untwtheory} .

First, the anomalous dimension of the lowest twisted mode,
$\eta_{1/2}=\Dh_{1/2}-\frac{1}{2}-\Delta_\phi$, is $-0.09$, $-0.11$, $-0.12$, $-0.13$, $-0.16$ for the reflection, Ising, $R_z(\pi)$, O(2) and the O(3)
centre twist (see Fig.~\ref{fig:physics}(left)), respectively. This matches the ordering of the one-loop attraction (see Eq. \eqref{eq:eqmu}):
$\mu_{\rm tw}\propto(n_{\rm tw}+2)/(M+8)=3/11<3/9<4/11<4/10<5/11$, and the one-loop values $-0.087,-0.101,-0.110,-0.119,-0.133$ are
   close to the data, with the largest differences modestly exceeding the quoted fit errors. Here the reference values subtract the same three-dimensional bulk $\Delta_\phi$ as the numerical data; a strictly first-order $\epsilon$-expansion would use $\Delta_\phi=1-\epsilon/2$ . At $s=3/2$ the anomalous part is reduced to $|\eta_{3/2}|\le0.08$ for all five defects, in line with the $1/|s|$ suppression of Eq.~\eqref{eq:oneloop1} and consistent with the one-loop values within $1$–$2.5\sigma$; the predicted spread across defects ($0.015$) is, however, smaller than the uncertainty of the $\psi_{3/2}$ extrapolations, so the ordering is tested only at $s=1/2$. Additionally, the bulk-to-defect couplings also follow the one-loop ordering nicely, $b_{\phi\psi_{1/2}}=0.841,0.857,0.859,0.876,0.880$ for the reflection, Ising, $R_z(\pi)$, O(2) and the centre twist,
i.e.\ $5$--$10\%$ above the free values and increasing with the attraction. The twisted single-mode sector    therefore shows a common qualitative pattern across the five defects. A generated zero-mode mass would feed back into the twisted potential~\cite{MCheng2026a}; we do not attempt to estimate this feedback from the three-dimensional spectra. The observed similarity  of the twisted spectra is an empirical result of the present no-explicit-pinning construction.

Second, Fig.~\ref{fig:physics} (right) compares the
interaction energies of all identified two-mode operators with the one-loop contact term. Antisymmetric channels (O(2) $\mathcal A$,
O(3) $A$, $[\hat\phi\partial\hat\phi]_A$) have interaction energies consistent with zero within errors. Symmetric channels are repulsive,
ordered by their flavour weights, and $1.1$--$1.8$ times stronger than at one loop in the channels compared here. 
The sign and flavour dependence of the contact interaction are thus robust features of the data.

Third, the converged defect-defect OPE coefficients of twisted modes lie within $7\%$ of their GFF
values: non-interacting channels are GFF-like and identical-mode channels are compatible with $\sqrt2$ or slightly below it (Ising $1.34(4)$, O(2) $1.39(3)$, O(3) $1.40$--$1.41$). The exceptions are the composites of untwisted $s=0$ modes, depleted by $9$--$25\%$, in the channel where perturbation theory breaks down (App.~\ref{sec:oneloop}) and where the generated defect mass flows (App.~\ref{sec:untwtheory}).


\section{Summary and discussion}
\label{sec:summary}

We have introduced a fuzzy-sphere realization of monodromy line defects in 3D CFTs: 
by inserting an Aharonov--Bohm $\pi$-flux through the poles on the two-sphere the symmetry-odd fermion  experiences a half-shifted angular-momentum grid.
This easy-to-implement scheme allows us to access various kinds of monodromy defects, including $\mathbb{Z}_2$ twist of the 3D Ising CFT, $\pi$-twist of the 3D O(2) CFT and three inequivalent monodromies of O(3) (centre twist,
$R_z(\pi)$ and the reflection $\phi_3\to-\phi_3$),  and obtain their low-lying spectra and OPE data.  
This yields a broad set of non-perturbative defect data, including scaling dimensions and bulk-to-defect and defect-defect OPE coefficients. Conceptually, the construction promotes the symmetry flux to a tunable parameter of the regularization, on the same footing as the pinning fields and impurities realized so far, and thereby gives access to  a class of conformal line defects that cannot be reached by any local deformation of the bulk Hamiltonian.

The twisted sector of all defects admits a simple description: the twisted modes form, to a good approximation, a generalized free field dressed by a flavour-symmetric contact interaction. The lowest twisted mode and the bulk-to-defect couplings follow the ordering set by the one-loop attraction, and the anomalous parts of the higher modes decrease with transverse spin as predicted. Two-mode operators are repelled in symmetric flavour channels, and their OPE coefficients deviate from generalized free theory by a few percent. 
The one-loop formulas derived here capture the sign and the flavour hierarchy of the two-mode shifts and the ordering of the single-mode towers, while they systematically underestimate the absolute repulsion.
This picture is reminiscent of the universality found at large transverse spin~\cite{Lemos:2017vnx}, but here it holds already for the lowest modes $s=\frac{1}{2}$, which suggests that monodromy defects of the O($N$) CFTs are a natural target for analytic bootstrap and large-spin methods, with the leading contact interaction providing a useful organizing principle rather than a complete description of the interacting defect.

For the partial twists, the absence of an explicit microscopic pinning term must be distinguished from the absence of a generated continuum defect mass. The same tadpole that controls the twisted spectra generates a mass for the untwisted zero mode, whose flow leaves the weak-coupling regime near four dimensions~\cite{MCheng2026a} (App.~\ref{sec:untwtheory}). The three-dimensional endpoint is then a non-perturbative question, addressed directly by the spectra and OPE data. In three dimensions, our enhanced zero-mode correlations and the absence of a resolved relevant invariant scalar support an interacting, symmetry-preserving conformal interpretation. They do not establish the ultimate endpoint solely from weak-coupling RG, nor exclude additional crossover effects beyond the fitted size dependence.  Ref.~\cite{MCheng2026a}  
complements this fixed-core construction by varying the quadratic mass, including zero explicit pinning. Establishing a common infrared endpoint would require control of core matching and finite-size effects. A future pole-coupling scan can test this, but no such scan or tuning is used in the results reported here.

Our results open a new avenue for the non-perturbative study of monodromy defects in three-dimensional CFTs. They also provide a framework for future calculations of defect free energies~\cite{Cuomo:2021rkm} and monodromy-pinning flows~\cite{Kravchuk:2025mpd}. 
The construction itself generalizes in several ways. Continuous monodromies $e^{2\pi ivQ}$ of U(1) symmetries are realized by fractional fluxes. Assuming particle--vortex duality, these monodromies provide the conformal data of fractional-charge Wilson lines in the critical Abelian Higgs model~\cite{Peskin:1977kp,Dasgupta:1981zz,Karch:2016sxi,Seiberg:2016gmd}, which we will present in a companion paper~\cite{Tang:2026xny}. And non-Abelian monodromies with continuous parameters, such as $e^{2\pi ivS^z}$ in O(3), follow in the same way. Applied to several copies of a CFT, the twist that cyclically permutes the replicas is the twist operator of the Renyi entropies~\cite{Hung:2014npa,Bianchi:2015liz}, so the same construction gives a direct non-perturbative route to the Renyi entropies and their universal corner and defect contributions in 3D CFTs. Finally, gauging the twisted symmetry turns the monodromy defects studied here into the vortex lines of $\mathbb Z_2$ and U(1) gauge theories, whose genuine defect spectra are given by the $g$-even sector of the present data, while the $g$-odd operators become endpoints of Wilson lines.

\section*{Data availability}
The numerical data (energies and matrix elements for all sizes and defects), the extrapolation and fit scripts, and the code used
to generate all figures and tables will be made publicly available in an open repository upon publication.

\section*{acknowledgments}
We thank M. Cheng for stimulating discussion and sharing  their unpublished draft with us.
W.Z. is supported by NSFC under No. 12625407, 12474144.  We used Claude (Anthropic, Opus 5.5 and Fable 5.1) to assist with drafting the manuscript, checking our results and writing the numerical codes under our direction. All AI-assisted content was reviewed and verified by the authors.

\vspace{0.5cm}

\textit{Note added.---} During the preparation of this manuscript, we became aware of another work~\cite{Qian2026}, which studies the Ising and O(2) monodromy defects on the fuzzy sphere by fixing the couplings by optimizing the energy scaling, whereas in this work, we keep the parameters at the bulk critical point. The partial monodromy defects of the O($N$) CFTs have also been studied in Ref.~\cite{MCheng2026a,Joydeep,MCheng2026b} by field theory and numerics.

\clearpage
\appendix

\section{One-loop analysis for arbitrary partial twists}
\label{sec:oneloop}
The derivation follows the logic of Refs.~\cite{Gaiotto:2013nva,Giombi:2021uae}: the quartic coupling produces (i) a one-body potential from the tadpole of the twisted fields and (ii) a contact interaction between defect modes. All quantities in this appendix are evaluated on the free monodromy background, with the standard (regular) boundary condition for every twisted mode and the free boundary condition for the untwisted $s=0$ modes. For a full twist   , perturbation theory about the regular free defect gives the fixed-point data to  this order. For partial twists the untwisted $s=0$ channel requires a separate treatment, which is discussed in App.~\ref{sec:untwtheory}.

\subsection{Setup and free monodromy defect}
We consider $M$ real scalars in $d=4-\epsilon$ with
\begin{gather}
S=\int d^dx\Big[\tfrac12(\partial\phi_a)^2+\tfrac{\lambda}{4!}(\phi_a\phi_a)^2\Big],\\
\frac{\lambda_*}{16\pi^2}=\frac{3\epsilon}{M+8}+O(\epsilon^2),
\label{eq:WF}
\end{gather}
at the Wilson--Fisher fixed point~\cite{Wilson:1971dc,Pelissetto:2000ek}, and a codimension-two monodromy defect that twists $n_{\rm tw}$ of the components ($g_a=-1$ for $a\le n_{\rm tw}$, $g_a=+1$
otherwise). At $\lambda=0$ each component decomposes into transverse partial waves $e^{is\varphi}$, with $s\in\mathbb Z+\frac{1}{2}$
for twisted and $s\in\mathbb Z$ for untwisted components. The defect operator with transverse spin $s$ corresponds to the
regular solution $\propto r^{|s|}$ of the transverse Laplace equation and has dimension
$\Dh_s=\frac{d-2}{2}+|s|$. The monodromy defect induces a nonzero coincident two-point function for the twisted components. Equivalently, the bilinear \(\phi_a^2\) acquires a nonzero one-point function in the presence of the defect. With the normalization
$\langle\phi_a(x)\phi_b(0)\rangle=\delta_{ab}/(4\pi^2x^2)$ in $d=4$, a    canonically normalized complex scalar $\Phi=(\phi_1+i\phi_2)/\sqrt2$  with monodromy $e^{2\pi iv}$ has
$\langle\bar\Phi\Phi\rangle=-v(1-v)/(8\pi^2r^2)$~\cite{Giombi:2021uae,Bianchi:2021snj}, so that each twisted real
component ($v=\frac{1}{2}$) has  the following bulk-vacuum-subtracted coincident contraction:
 \begin{gather}
\langle\phi_a(x)\phi_b(x)\rangle_{\mathcal D}=\delta_{ab}\,\frac{C_a}{r^2},\\
C_a=-\frac{1}{32\pi^2}\ \ (g_a=-1),\qquad C_a=0\ \ (g_a=+1).
\end{gather}

\subsection{Induced potential}
At first order in $\lambda$ the quartic vertex with one pair of legs contracted generates a quadratic, position-dependent
term. Contracting one pair in $(\phi_c\phi_c)(\phi_d\phi_d)$ in all possible ways gives
$2\langle\phi_c\phi_c\rangle\,\phi_d\phi_d+4\langle\phi_a\phi_b\rangle\phi_a\phi_b$, so that
\begin{equation}
\begin{aligned}
\frac{\lambda}{4!}(\phi\cdot\phi)^2\ &\to\ \sum_a\frac{\mu_a}{2r^2}\,\phi_a^2,\\
\mu_a=\frac{\lambda}{6}\Big(\sum_cC_c+2C_a\Big)&=\frac{\lambda C}{6}\,\big(n_{\rm tw}+2\,\delta_{g_a,-1}\big),
\end{aligned}
\label{eq:mua}
\end{equation}
with $C=-1/(32\pi^2)$. At the fixed point~\eqref{eq:WF},
\begin{equation}
\mu_{\rm tw}=-\frac{(n_{\rm tw}+2)\,\epsilon}{4(M+8)},\qquad \mu_{\rm untw}=-\frac{n_{\rm tw}\,\epsilon}{4(M+8)}.
\label{eq:mu}
\end{equation}
Both potentials are attractive. The twisted components feel the larger one because the self-contraction of a field contributes in addition to the trace. These are the potentials on the free background; a non-trivial boundary condition of the untwisted $s=0$ modes would modify both~\cite{MCheng2026a}.

\subsection{Single-mode dimensions}
A potential $\mu/(2r^2)$ modifies the transverse equation of a partial wave to
$[-\partial_r^2-\frac{1}{r}\partial_r+(s^2+\mu)/r^2]f=0$, whose    standard-branch  solution is $r^{\nu}$ with $\nu=\sqrt{s^2+\mu}$.    In this inverse-square approximation,
 \begin{equation}
\begin{aligned}
\Dh_s&=\frac{d-2}{2}+\sqrt{s^2+\mu}\\
&=1-\frac{\epsilon}{2}+|s|+\frac{\mu}{2|s|}+O(\epsilon^2)\qquad(s\neq0).
\end{aligned}
\label{eq:oneloop1}
\end{equation}
The same result follows from first-order perturbation theory on the cylinder $\mathbb R\times S^3$: there the defect wraps
a great circle, the potential becomes $\mu/(2\sin^2\chi)\,\phi^2$ with $\chi$ the angle from the defect, and the
single-mode state has wavefunction $Y_s\propto(\sin\chi)^{|s|}e^{is\varphi}$ and energy $\Delta=1+|s|$; the expectation value of
the potential is $\mu\int|Y_s|^2/\sin^2\chi\,/(2\Delta)=\mu/(2|s|)$ (App.~\ref{app:s3}). For $n_{\rm tw}=M$,
Eq.~\eqref{eq:oneloop1} reproduces Refs.~\cite{Gaiotto:2013nva,Soderberg:2017oaa,Giombi:2021uae}; e.g.\ for Ising
$\Dh_{1/2}=\frac{3}{2}-\frac{7\epsilon}{12}$. The square root resums repeated insertions of this particular potential and should not be read as an all-order result; the expansion requires $|\mu|\ll s^2$ and therefore excludes the untwisted $s=0$ modes.

\subsection{Two-mode operators}
On the cylinder the quartic interaction $H_{\rm int}=\frac{\lambda}{4!}\int_{S^3}(\phi\cdot\phi)^2$ shifts the energy of a
two-particle state $|\psi_{a,s_1}\psi_{b,s_2}\rangle$. With the canonical normalization of a field quantum of energy $\Delta$,
$\phi\supset(2\Delta)^{-1/2}(a\,Y+a^\dagger\bar Y)$, the first-order shift for two distinct modes of a single real field is
\begin{equation}
\begin{aligned}
\delta E&=\frac{\lambda}{4!}\cdot4!\int_{S^3}\frac{|Y_{s_1}|^2|Y_{s_2}|^2}{2\Delta_1\,2\Delta_2}\\
&=\frac{\lambda}{8\pi^2}\,\frac{1}{1+|s_1|+|s_2|},
\end{aligned}
\label{eq:dE}
\end{equation}
where the overlap integral is evaluated in App.~\ref{app:s3}; the normalization factors $(1+|s_i|)$ of the
wavefunctions cancel against $2\Delta_i$, leaving only the transverse overlap. For identical modes the two-particle state
is $(a^\dagger)^2|0\rangle/\sqrt2$ and the combinatorics give half of Eq.~\eqref{eq:dE}. For general flavours, the vertex
$(\phi\cdot\phi)^2$ acts on the flavour state of the pair through the symmetric tensor
$\frac{1}{3}(\delta_{ab}\delta_{cd}+\delta_{ac}\delta_{bd}+\delta_{ad}\delta_{bc})$, whose eigenvalue $T_R$ multiplies Eq.~\eqref{eq:dE}. Inserting
$\lambda_*$, we obtain
\begin{equation}
\begin{aligned}
\gamma^{\rm c}_{s_1s_2}={}&\frac{3\epsilon}{M+8}\;T_R\;\frac{2}{1+|s_1|+|s_2|}\\
&\times\begin{cases}1 & s_1\neq s_2,\\ \tfrac12 & s_1=s_2,\end{cases}
\end{aligned}
\label{eq:oneloop2}
\end{equation}
with the flavour weights
\begin{equation}
T_R=\begin{cases}\frac{M'+2}{3} & \text{singlet of O}(M'),\\ \frac{2}{3} & \text{symmetric traceless},\\ 0 & \text{antisymmetric},\\ \frac{1}{3} & \text{one twisted and one untwisted mode},\end{cases}
\end{equation}
where O($M'$) is the symmetry acting on the two modes (O($n_{\rm tw}$) for twisted pairs, O($M-n_{\rm tw}$) for untwisted
pairs). Explicitly, the vertex acts on a two-particle flavour state as $S|ab\rangle=\tfrac13\big(\delta_{ab}\sum_c|cc\rangle+|ab\rangle+|ba\rangle\big)$: for $a\neq b$
the symmetric and antisymmetric combinations $|ab\rangle\pm|ba\rangle$ give $\frac{1}{3}\pm\frac{1}{3}$, i.e.\ $\frac{2}{3}$ and $0$; and the O($M'$)
singlet $\sum_a|aa\rangle/\sqrt{M'}$ gives $(M'+2)/3$. In the complex O(2) basis $\Psi=(\psi^1+i\psi^2)/\sqrt2$ the $C$-even and $C$-odd
combinations of $\Psi\bar\Psi$ are the singlet and the antisymmetric tensor, and $\Psi\Psi$ is the symmetric traceless (charge-2)
tensor, so the same weights apply.    For a nondegenerate pair that defines a primary, its first-order dimension  is $\Dh_{[12]}=\Dh_{s_1}+\Dh_{s_2}+\gamma^{\rm c}_{s_1s_2}$.  For a partial twist this is a reference-background expression until the zero-mode sector has been specified consistently.  Two-mode primaries that are odd under
the exchange of their constituents, such as $\psi\partial_\tau\bar\psi-(\partial_\tau\psi)\bar\psi$ (the Ising $p^o$), vanish at coincident points
and receive no contact contribution at this order, $\Dh=\Dh_{s_1}+\Dh_{s_2}+1$.

For a twisted--untwisted pair the two constituents live on different spin grids, so only the direct term of the vertex acts within the allowed two-particle space; this is why $T=\frac{1}{3}$ rather than the symmetric-traceless value $\frac{2}{3}$. Equation~\eqref{eq:oneloop2} is a diagonal matrix element and gives a dimension only if the pair state is not degenerate with another state of the same quantum numbers at the free level. Otherwise the full first-order mixing matrix, including the different single-mode shifts, has to be diagonalized~\cite{Gaiotto:2013nva}; examples are $[\psi_{1/2}\psi_{5/2}]$ and $[\psi_{3/2}\psi_{3/2}]$ at $L_z=3$ for Ising, and, for the partial twists, twisted and untwisted pairs in the same symmetry sector, such as $[\psi^{(3)}_{1/2}\psi^{(3)}_{1/2}]$ and $[\hat\phi^{(0)}\hat\phi^{(1)\dagger}]$ at $L_z=1$ for the reflection (App.~\ref{app:fits}).

\subsection{Consistency checks}
(i) For Ising ($M=n_{\rm tw}=1$, $T=1$) the singlet $s=[\psi_{1/2}\bar\psi_{1/2}]$ gets $\gamma=\epsilon/3$, hence
$\Dh_s=3-\frac{5\epsilon}{6}$, and $t_+=[\psi_{1/2}\psi_{3/2}]$ gets $\Dh=4-\frac{8\epsilon}{9}$, in agreement with
Ref.~\cite{Gaiotto:2013nva}. (ii)    For a full twist ($n_{\rm tw}=M$), the  displacement operator is the O(   $M$) singlet  of two identical modes,    $\hat D\propto\sum_a\psi_{a,1/2}\psi_{a,1/2}$. With $T=(M+2)/3$ and the identical-mode factor $1/2$, the contact shift $(M+2)\epsilon/[2(M+8)]$ cancels $2\mu_{\rm tw}$  in $2\Dh_{1/2}$, giving $\Dh_{\hat D}=3-\epsilon=d-1$   . This independently checks the normalization and combinatorics  of Eq.~\eqref{eq:oneloop2}  . For a partial twist    the same algebra cancels the twisted diagonal contribution, but  the twisted singlet is degenerate    with an untwisted singlet pair at total transverse spin one. It therefore does not alone identify the displacement eigenoperator. The protected dimension follows from the conformal Ward identity at an actual defect  fixed point,    after the  zero-mode boundary condition    and allowed mixing have been included . All one-loop values quoted
below are evaluated at $\epsilon=1$.

\subsection{Large transverse spin}
At large  positive  transverse spin,    $\Dh_s=s+\Delta_\phi+c_{\rm ls}\,s^{-\tau_{\rm min}/2}+\dots$ ~\cite{Lemos:2017vnx}; for Ising this gives the
``LLMS'' values in Table~\ref{tab:ising}.

\subsection{The untwisted zero mode with no explicit microscopic pinning}
\label{sec:untwtheory}

All numerical results in this paper use the bulk critical Hamiltonian with the orbital sets changed only to impose the monodromy. No independent local term is added or tuned at the north or south pole. We call this the \emph{no-explicit-pinning} or \emph{unpinned} construction. This specifies the microscopic Hamiltonian; continuum defect couplings remain subject to renormalization. The pole regularization fixes their ultraviolet matching conditions, including finite contributions to the allowed defect couplings.

Let $M_{\rm u}=M-n_{\rm tw}>0$ denote the number of untwisted components and $p=d-2$ the defect dimension. At the free reference defect, the untwisted $s=0$ field has dimension $p/2$, so its invariant quadratic composite is classically marginal. The continuum theory therefore allows
\begin{equation}
S_{\mathcal D}=\frac{c}{2}\int d^p y\,\hat\phi_0^2(y),
\qquad
\hat\phi_0^2\equiv
\sum_{i\in\mathrm{untw}}\hat\phi_{i,0}^2.
\label{eq:Sdef}
\end{equation}
This coupling preserves the defect symmetries. It must be distinguished from a linear source for $\hat\phi_0$ or $\hat\psi_{1/2}$, which explicitly breaks some of those symmetries. Consequently, a generated quadratic mass does not itself imply symmetry breaking, although its subsequent flow can affect the infrared phase.

The renormalization of these defect masses was studied in Ref.~\cite{MCheng2026a}. For a partial monodromy, bulk interactions generate an untwisted mass even when its initial value vanishes. Near four dimensions, the induced flow favors a negative mass and enhances untwisted correlations. Thus the regular free zero-mode condition is not preserved by renormalization. We use this result to interpret the unpinned construction: eliminating an explicit pole potential does not eliminate the defect flow, but selects a microscopic starting point for it. The sign of the generated interaction provides a physical explanation for enhanced zero-mode correlations without introducing negative pinning by hand.

Ref.~\cite{MCheng2026a} also identifies regimes near four dimensions in which the untwisted mass runs out of weak coupling, with no perturbative fixed point controlling the infrared behavior. This runaway establishes the breakdown of perturbation theory along that trajectory, but does not determine its endpoint in three dimensions. 
Our calculations address this question directly at the accessible sizes. 

Our spectra are compatible with the conformal interpretation at accessible sizes. For the O(3) rotational defect $R_z(\pi)$ and the reflection defect, respectively, we obtain untwisted dimensions $\hat\Delta_0=0.46(2)$ and $0.487(8)$, below the bulk order-parameter dimension. The corresponding resolved invariant singlets have dimensions $1.26(4)$ and $1.452(5)$, above the line-defect marginality threshold of one. Enhanced fundamental correlations and an irrelevant invariant mass are compatible because the interacting composite need not have dimension $2\hat\Delta_0$. These observations support local stability in a symmetric mass direction, provided the measured singlet is identified with the corresponding scaling operator. They do not establish stability against every allowed perturbation, and operator mixing must be considered when making that identification.

The Monte Carlo results of Ref.~\cite{MCheng2026a} provide a useful comparison. Without explicit pinning, that work reports untwisted dimensions $0.409(11)$ for the rotational defect and $0.472(6)$ for the reflection.
These are fitted exponents at the simulated sizes, and the reference discusses their drift with pinning strength. It also reports evidence for discrete symmetry breaking at sufficiently strong rotational pinning. The comparison shows that untwisted correlations can respond strongly to the microscopic core coupling, especially for the rotational defect. Our estimates exhibit the same qualitative enhancement, although our rotational estimate is appreciably larger. These differences alone do not establish distinct infrared universality classes. Different core matching conditions and accessible length scales can produce different crossover behavior. Conversely, qualitative agreement alone does not establish a common endpoint: that requires checking the asymptotic spectra and their dependence on the core regularization. Systematic comparisons across sizes are therefore essential.

This perspective also clarifies the perturbative comparison in Tables~\ref{tab:o3r} and~\ref{tab:o3z}. The free value $\hat\Delta_0=1/2$ used there is a reference prescription for organizing levels together with the contact interaction shifts. It is not a controlled prediction for the interacting untwisted field or for composites containing it. Our numerical energies, extrapolations, and OPE coefficients are obtained independently of this prescription. Likewise, reference-background corrections to nonzero-spin modes should not automatically be identified with corrections at the eventual interacting defect fixed point. Such comparisons depend on the background about which the expansion is performed and cannot determine an unknown three-dimensional running mass from the measured spectrum alone.

The physical question is therefore whether the microscopic construction reaches a symmetry-preserving defect CFT, and whether different core realizations approach the same theory. No explicit pinning and weak symmetric pinning can share an endpoint when their matched couplings lie in the same infrared basin, but monodromy alone does not guarantee this. The present spectra support a conformal interpretation without resolving every possible long-distance crossover. In particular, the near-four-dimensional runaway does not invalidate the numerical evidence in three dimensions, just as the large-$N$ conformal solution does not prove the endpoint for O(3). The fit uncertainties in App.~\ref{app:fits} retain their stated scope and do not by themselves quantify this additional dynamical question.

A direct extension would vary symmetry-preserving pole operators and follow the resulting spectra and OPE coefficients with increasing system size. Agreement across a range of core couplings would test universality, while persistent changes could reveal crossover effects or distinct defect phases. A microscopic pole operator generally overlaps with several defect singlets, so identifying its projection onto the mass scaling field is part of this analysis. Such a scan would connect our fixed unpinned realization to the tunable mass deformations studied in Ref.~\cite{MCheng2026a}. All data reported here retain the original bulk Hamiltonian and impose only the monodromy boundary condition.

\section{Mode functions and overlaps on $S^3$}
\label{app:s3}
In $d=4$ we map $\mathbb R^4$ to $\mathbb R\times S^3$. With the defect spanning $(x_1,x_2)$, we parametrize the unit $S^3$ by
$x_1+ix_2=\cos\chi\,e^{i\tau'}$, $x_3+ix_4=\sin\chi\,e^{i\varphi}$, $\chi\in[0,\frac{\pi}{2}]$, so that the defect intersects $S^3$ along the
great circle $\chi=0$ and $d\Omega_3=\sin\chi\cos\chi\,d\chi\,d\varphi\,d\tau'$ (total volume $2\pi^2$). The defect primary of transverse
spin $s$ is created by the homogeneous function $(x_3+ix_4)^{|s|}$ (or its conjugate), which for non-integer $|s|$ is locally
harmonic on $\mathbb R^4\setminus\mathcal D$ on a chosen branch, with the prescribed monodromy around the defect; this is precisely the
twisted mode. On $S^3$ its wavefunction is
\begin{equation}
Y_s=\mathcal N_s(\sin\chi)^{|s|}e^{is\varphi},\qquad \mathcal N_s^2=\frac{|s|+1}{2\pi^2},
\end{equation}
with energy $\Delta_s=1+|s|$ on the unit-radius cylinder. The integrals needed in App.~\ref{sec:oneloop} are
\begin{equation}
\begin{aligned}
\int_{S^3}\frac{|Y_s|^2}{\sin^2\chi}&=\frac{|s|+1}{|s|},\\
\int_{S^3}|Y_{s_1}|^2|Y_{s_2}|^2&=\frac{(|s_1|+1)(|s_2|+1)}{2\pi^2\,(1+|s_1|+|s_2|)} .
\end{aligned}
\end{equation}
With the canonical normalization $(2\Delta)^{-1/2}$ of a field quantum, the single-mode and two-mode shifts are
\begin{align}
&\mu\,\frac{|s|+1}{|s|}\,\frac{1}{2(1+|s|)}=\frac{\mu}{2|s|},\\
&\lambda\,\frac{(|s_1|+1)(|s_2|+1)}{2\pi^2(1+|s_1|+|s_2|)}\,\frac{1}{4(1+|s_1|)(1+|s_2|)}\nonumber\\
&\qquad=\frac{\lambda}{8\pi^2(1+|s_1|+|s_2|)},
\end{align}
which are Eqs.~\eqref{eq:oneloop1} and~\eqref{eq:dE}. The divergence of the single-mode shift at $s=0$ signals that the potential cannot
be treated as a small perturbation in the untwisted $s=0$ channel. The formal inverse-square solution gives \(\nu=\sqrt{\mu_{\rm untw}}\), which is imaginary at this order. Determining the physical scaling dimension therefore requires a treatment beyond this naive one-loop approximation, including the appropriate defect-local coupling or boundary condition.    The corresponding weak-coupling flow is discussed in App.~\ref{sec:untwtheory}; it does not determine the three-dimensional endpoint, which is why the free value $\frac{1}{2}$ used for this mode in Tables~\ref{tab:o3r} and~\ref{tab:o3z} is a reference value rather than a fixed-point prediction.

\section{Matrix elements with twisted orbitals}
\label{app:me}
The microscopic Hamiltonian is not modified by the defect: the real-space interaction is a fixed combination of $\delta(\mathbf x)$ and
$\nabla^2\delta(\mathbf x)$ whose two coefficients are determined once, at each $\Norb$, by the bulk pseudopotentials $V_0,V_1$ of the
untwisted model. Every projected two-body matrix element and every self-contraction term of the defect Hamiltonian is then
recomputed from this continuum interaction with the actual orbital sets~\eqref{eq:Malpha} of the twisted and untwisted flavours; no
bulk pseudopotential table is reused for a twisted flavour.
With $u=\cos^2\frac{\theta}{2}$, $w=\sin^2\frac{\theta}{2}$ and the normalized orbitals of Eq.~\eqref{eq:orb} written as
$Y^{(s)}_m=\mathcal N_m e^{im\varphi}u^{\alpha_m}w^{\beta_m}$, $\alpha_m=\frac{s+m}{2}$, $\beta_m=\frac{s-m}{2}$, valid for integer and half-integer
$m$, all integrals reduce to $\int_{-1}^{1}u^pw^q\,d\cos\theta=2B(p+1,q+1)$, with $\mathcal N_m^{-2}=4\pi B(2\alpha_m+1,2\beta_m+1)$. For the contact
interaction
\begin{equation}
\begin{aligned}
&\int d\Omega\,\bar Y_1\bar Y_2Y_3Y_4\\
&\quad=2\pi\,\delta_{m_1+m_2,m_3+m_4}\prod_i\mathcal N_i\;2B\Big(\sum_i\alpha_i+1,\sum_i\beta_i+1\Big),
\end{aligned}
\end{equation}
where $Y_i\equiv Y^{(s)}_{m_i}$, and for $\nabla^2\delta$ we use, on the unit sphere,
\begin{equation}
\begin{aligned}
\nabla^2\big(u^Aw^Be^{ik\varphi}\big)={}&\Big[A^2u^{A-1}w^{B+1}+B^2u^{A+1}w^{B-1}\\
&-(2AB+A+B)u^Aw^B\\
&-\tfrac{k^2}{4}u^{A-1}w^{B-1}\Big]e^{ik\varphi}.
\end{aligned}
\end{equation}
Physical units follow from $\delta(\mathbf x)=\delta(\hat{\mathbf n})/R^2$ and $\nabla^2=\nabla^2_{\hat{\mathbf n}}/R^2$. 
Since we implement
Eq.~\eqref{eq:Hguo} without normal ordering, following the convention of Ref.~\cite{Guo:2025onf}, the self-contractions of the interaction
generate one-body terms,
\begin{equation}
\begin{aligned}
&\frac{1}{R^2}\sum_{m\in\mathcal M_\alpha}\!\int\!\bar Y_{m_1}Y_{m_1}|Y_m|^2\quad\text{and}\\
&-\frac{U_1}{R^4}\sum_\beta\sum_a|\tau^a_{\alpha\beta}|^2\sum_{m\in\mathcal M_\beta}\int\bar Y_{m_1}Y_m\nabla^2(\bar Y_mY_{m_1}),
\end{aligned}
\end{equation}
with each sum over the orbital set of the contracted flavour. In the defect these terms therefore also use twisted
orbitals. All formulas hold for both grids ($m=-s,\dots,s$ and $m=-s+\frac{1}{2},\dots,s-\frac{1}{2}$), since they depend on $m$ only through the
exponents $\alpha_m,\beta_m$; for mixed twisted/untwisted quadruples they were checked against direct numerical quadrature to
$10^{-12}$. No additional one-body or two-body term is placed at the poles: the defect is defined solely by the orbital sets~\eqref{eq:Malpha}. In particular, no explicit mass pinning of the untwisted flavours is included,    whereas  Ref.~\cite{MCheng2026a}  additionally varies a lattice mass term. Its zero-explicit-pinning case is the appropriate first comparison to our fixed microscopic prescription .

\section{Symmetries of monodromy defects in O(3) CFT}
\label{app:o3sym}
In the flavour basis $(\psi_+,\psi_z,\psi_-,\psi_0)$ of Sec.~\ref{sec:bases} all three O(3) monodromies are diagonal, and the twist
of every CFT operator is fixed by that of its constituent fermions: a bilinear $\psi^\dagger_\alpha\psi_\beta$ is antiperiodic around the
defect if exactly one of $\alpha,\beta$ is twisted, and periodic otherwise.
The monodromy of the CFT operators follows from the rule stated above and is summarized in
Table~\ref{tab:o3ops}. For the centre twist the whole vector is antiperiodic; for $R_z(\pi)$ the component
$\phi_3$ stays periodic and yields integer-spin modes $\hat\phi_3^{(s)}$; for the reflection the roles are
exchanged: $\phi_3$ is antiperiodic (modes $\psi^{(3)}_s$, $s\in\mathbb Z+\frac{1}{2}$) and the doublet $(\phi_1,\phi_2)$
stays periodic (modes $\hat\phi^{(s)}$, $s\in\mathbb Z$, with U(1) charge $\pm1$). 
By the bulk-to-defect expansion~\eqref{eq:bdexp}, an antiperiodic operator couples only to defect operators with half-integer
transverse spin and a periodic one only to integer-spin defect operators. The table therefore fixes which bulk-to-defect couplings
can be nonzero for each defect. The singlet $S$ and the tensors $t_{++}$ and $t_{33}$ are periodic for all three twists    and can couple to symmetry-compatible integer-spin defect operators. Only components invariant under the preserved internal symmetry  can acquire one-point functions   : $t_{++}$ is charged and has zero one-point function, while $t_{33}$ can have a nonzero one-point function for the partial twists ; $t_{33}$ is the probe used for the Ward identity in
App.~\ref{app:O3}. The charge-1 tensor $t_{+3}$, in contrast, is antiperiodic for $R_z(\pi)$ and the reflection, so its leading
defect operators carry $L_z=\pm\frac{1}{2}$ and are mixed operators of one twisted and one untwisted mode.

\begin{table}[t]
\centering\footnotesize
\caption{Monodromy of the O(3) CFT operators under the three twists, as inherited from the fermion bilinears.
A: antiperiodic (half-integer transverse spin); P: periodic (integer transverse spin).
}
\label{tab:o3ops}
\setlength{\tabcolsep}{3.5pt}
\begin{tabular}{llccc}
\toprule
operator & fermion bilinear & centre twist & $R_z(\pi)$ & reflection \\
\midrule
$\phi_\pm$ & $\psi_0^\dagger\psi_\pm+\psi_\mp^\dagger\psi_0$ & A & A & P \\
$\phi_3$ & $\psi_0^\dagger\psi_z+\mathrm{h.c.}$ & A & P & A \\
$t_{++}$ (charge 2) & $\psi_+^\dagger\psi_-$ & P & P & P \\
$t_{+3}$ (charge 1) & $\psi_+^\dagger\psi_z+\psi_z^\dagger\psi_-$ & P & A & A \\
$t_{33}$ & $\psi_z^\dagger\psi_z-\tfrac13\sum_a\psi_a^\dagger\psi_a$ & P & P & P \\
$S$ & $\psi_0^\dagger\psi_0$ & P & P & P \\
\bottomrule
\end{tabular}
\end{table}

\section{Bulk-to-defect couplings and OPE coefficients of the O(3) defects}
\label{app:O3}
\begin{figure*}[t]
\centering
\includegraphics[width=\textwidth]{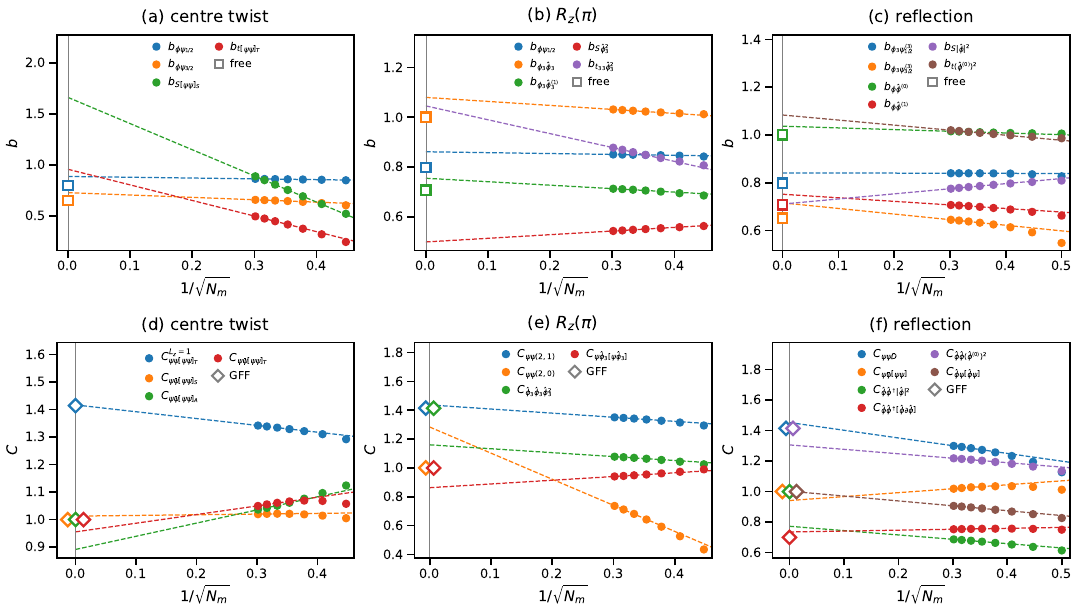}
\caption{Bulk-to-defect couplings (top row) and defect-defect OPE coefficients (bottom row) of the three O(3) monodromy defects:
(a,d) centre twist $g=-\one$, after the Clebsch--Gordan reduction; (b,e) $R_z(\pi)$; (c,f) reflection $\phi_3\to-\phi_3$. Dashed lines
are $1/\sqrt{\Norb}$ fits to the four largest sizes. Reference values at $1/\sqrt{\Norb}=0$, in the colour of the corresponding data:
squares for the free theory, diamonds for GFF.}
\label{fig:o3ope}
\end{figure*}

This appendix collects the bulk-to-defect couplings, one-point functions and defect-defect OPE coefficients of the three O(3)
monodromy defects whose spectra are discussed in Sec.~\ref{sec:o3}. They are obtained numerically using Eqs.~\eqref{eq:ab} and~\eqref{eq:C3};
Fig.~\ref{fig:o3ope} shows their finite-size scaling, and Tables~\ref{tab:o3c_ope}--\ref{tab:o3z_ope} list the values at the
largest size together with the extrapolations of App.~\ref{app:fits}. For the centre twist, matrix elements between $S^z$
components are divided by the Clebsch--Gordan coefficient $|\langle1,{+}1;1,{-}1|R,0\rangle|$ to obtain the coefficient of the
normalized O(3) multiplet $R$.

\paragraph{Ward identity and $C_D$.} The Ward identity of the displacement operator relates $a_{\mathcal O}$ to the bulk-to-defect
coupling $b_{\mathcal O\hat D}$ and to $C_D$~\cite{Billo:2016cpy}; in $d=3$ and with $\hat D$ unit normalized~\cite{Hu:2023ghk},
\begin{equation}
C_D=\Big[\frac{2\,\Delta_{\mathcal O}\,a_{\mathcal O}}{\pi\, b_{\mathcal O\hat D}}\Big]^2 .
\label{eq:CD}
\end{equation}
Here $C_D$ is the two-point coefficient of the complex displacement $D_\pm=D_x\pm iD_y$, with $\langle D_+(y)D_-(0)\rangle=C_D/|y|^4$ and $\hat D_\pm=D_\pm/\sqrt{C_D}$. In the Cartesian convention $\langle D_i(y)D_j(0)\rangle=C_D^{\rm cart}\delta_{ij}/|y|^4$, one has $C_D=2C_D^{\rm cart}$. This complex-spin convention  gives $C_D=1/(2\pi^2)$ per twisted real component for the free scalar. Since $C_D$ is a square, we also use
the unsquared Ward ratio $R_D\equiv2\Delta_{\mathcal O}a_{\mathcal O}/(\pi b_{\mathcal O\hat D})$, with $C_D=R_D^2$, and quote $|R_D|$ when
discussing its finite-size behaviour. We evaluate Eq.~\eqref{eq:CD} with the tensor $t_{33}$ (microscopically
$\psi_z^\dagger\psi_z-\frac{1}{3}\sum_a\psi_a^\dagger\psi_a$, the $m=0$ component of the O(3) symmetric traceless tensor), which, unlike the
singlet densities, cannot mix with the stress tensor and therefore gives a clean one-point function. This is possible for the
$R_z(\pi)$ and reflection defects, which preserve only O(2)$\times\mathbb Z_2$; for the centre twist, which preserves the full O(3),
$\langle t_{33}\rangle$ vanishes by symmetry.    Equation~\eqref{eq:CD} uses the conformal forms of the correlators. If a partial twist is still in a crossover regime, this extraction can have additional scale dependence. The drift reported below does not by itself distinguish such a flow from ordinary finite-size corrections  (App.~\ref{sec:untwtheory}).

\paragraph{Centre twist.} The coupling of the order parameter to the lowest mode,
$b_{\phi\psi_{1/2}}=0.880(6)$, is the largest of the five $\mathbb Z_2$ defects, $10\%$ above the free value $\sqrt{2/\pi}=0.798$, and
$b_{\phi\psi_{3/2}}=0.71(2)$ lies $9\%$ above its free value $0.651$ (Table~\ref{tab:o3c_ope}); the couplings of the bulk singlet $S$ and
tensor $t$ to the two-mode singlet and quintet are still drifting at these sizes.
After the Clebsch--Gordan reduction, the three $\psi\bar\psi$ channels give
$C_{\psi\bar\psi[\psi\psi]_S}=1.01(5)$, $C_{\psi\bar\psi[\psi\psi]_A}=0.93(4)$ and $C_{\psi\bar\psi[\psi\psi]_T}=0.98(3)$, all compatible with the GFF
value $1$, and the identical-mode channel gives $C_{\psi\psi[\psi\psi]_T^{L_z=1}}=1.40(4)$, compatible with $\sqrt2$ and, as for Ising and
O(2), slightly below it. The coupling of two different modes to the quintet, $C_{\psi_{1/2}\psi_{-1/2}[\psi\psi]_T}$, is not converged.

\paragraph{$R_z(\pi)$.} In the twisted sector, $b_{\phi\psi_{1/2}}=0.859(3)$ and $C_{\psi\psi(2,1)}=1.41(3)$ (Table~\ref{tab:o3r_ope}), close to
the O(2) values. At $\Norb=11$ the one-point estimate of the symmetric tensor,
which is free of bulk stress-tensor admixture,
is $a_{t_{33}}=0.179$, slightly below its free value $0.184$; the extrapolation gives $0.204(9)$, about $11\%$ above the free value. The Ward identity~\eqref{eq:CD} requires
$b_{t_{33}\hat D}$, which increases steadily with $\Norb$; the unsquared Ward ratio $|R_D|=2\Delta_t|a_{t_{33}}|/(\pi b_{t_{33}\hat D})$
decreases monotonically from $0.71$ ($\Norb=5$) to $0.35$ ($\Norb=11$); correspondingly $C_D=R_D^2$ decreases from $0.50$ to $0.12$,
towards the free value $1/\pi^2=0.10$. The sequence has not flattened, and its extrapolations depend strongly on the fit form; we
therefore report only the trend. In the untwisted channel, the coupling of $\phi_3$ to its $s=0$ mode, $b_{\phi_3\hat\phi_3}=1.07(1)$,
is close to the trivial-defect value $1$, and that to the $s=1$ mode, $b_{\phi_3\hat\phi_3^{(1)}}=0.74(1)$, is close to its free value
$1/\sqrt2$. The OPE coefficient $C_{\hat\phi_3\hat\phi_3\hat\phi_3^2}=1.14(7)$ of the lowest $\mathbb Z_2$-even singlet $\hat\phi_3^2$ is about
$19\%$ below the GFF value $\sqrt2$, the second largest deviation from GFF among the converged data, after $|\hat\phi|^2$ of the
reflection defect ($-25\%$), while the mixed operator $[\psi_{1/2}\hat\phi_3]$ has $C_{\psi\hat\phi_3[\psi\hat\phi_3]}=0.88(2)$.

\paragraph{Reflection.} The coupling of $\phi_3$ to the lowest twisted mode is $b_{\phi_3\psi^{(3)}_{1/2}}=0.841(3)$ (Table~\ref{tab:o3z_ope}).
Since only one component is twisted, the displacement is built from $\psi^{(3)}$ alone at the free level: $C_{\psi^{(3)}\psi^{(3)}\bar D}$
is $1.30$ at $\Norb=11$ and rises towards $\sqrt2$, while the coupling of the displacement to the untwisted modes,
$C_{\hat\phi^{(0)}\hat\phi^{(1)\dagger}\bar D}$, vanishes in free theory and is only $0.12$ at $\Norb=11$. The twisted two-mode singlet
$[\psi^{(3)}\bar\psi^{(3)}]$ has a GFF-like OPE coefficient $0.96(3)$; it is the analogue of the Ising singlet $s$, embedded in O(3).
The estimate $a_{t_{33}}=-0.181(3)$, which is free of bulk
stress-tensor admixture, has the \emph{opposite sign}
to that of the $R_z(\pi)$ defect and agrees with
the free value $-0.184$: for $R_z(\pi)$ the twisted pair $(\phi_1,\phi_2)$ is suppressed near the line and
$\phi_3$ is relatively enhanced, while here $\phi_3$ itself is suppressed.
The unsquared Ward ratio $|R_D|$ decreases from $0.92$ to $0.30$ over $\Norb=4$--$11$; correspondingly $C_D=R_D^2$ decreases from $0.85$ to
$0.09$, towards the free value $1/(2\pi^2)=0.051$, without having converged. In the untwisted doublet, $b_{\phi\hat\phi^{(1)}}=0.74(1)$ is
close to its free value $1/\sqrt2$ and $b_{\phi\hat\phi^{(0)}}=1.031(6)$ close to the trivial-defect value. Compared with $R_z(\pi)$,
the one-loop attraction in this channel is half as strong ($\mu_{\rm untw}=-1/44$ instead of $-2/44$), and both deviations from the
free untwisted mode are smaller: $b-1$ by a factor of $2$, and the downward shift of the extrapolated dimension from the trivial-defect value $\Delta_\phi$ by a factor of about 2 ($0.059$ vs $0.032$), matching the ratio of $\mu_{\rm untw}$. The OPE coefficients of the composites of $s=0$ modes are depleted relative to GFF:
$C_{\hat\phi\hat\phi^\dagger|\hat\phi|^2}=0.75(6)$ ($-25\%$) and $C_{\hat\phi\hat\phi(\hat\phi^{(0)})^2}=1.28(5)$ ($-9\%$), whereas channels involving a
mode with $s\neq0$ are GFF-like [$C_{\hat\phi\psi^{(3)}[\hat\phi\psi^{(3)}]}=0.98(5)$, $C_{\hat\phi\hat\phi^\dagger[\hat\phi\partial\hat\phi]_A}=0.740(6)$ vs.\
$\sqrt{\Dh_{\hat\phi^{(0)}}}\simeq0.70$]. The coefficients involving $\hat\phi^{(1)}$ or $(\hat\phi^{(0)})^3$ are not converged. This is a qualitative tendency and does not fix a universal ratio of the three-dimensional shifts: matching conditions and the subsequent RG evolution also enter.

\begin{table}[t]
\centering\footnotesize
\caption{O(3) centre twist: couplings, one-point functions and defect-defect OPE coefficients (O(3) multiplet coefficients, after Clebsch--Gordan reduction).}
\label{tab:o3c_ope}
\fitwidth{%
\begin{tabular}{lcccc}
\toprule
quantity & sizes & largest & FS & free / GFF \\
\midrule
$b_{\phi\psi_{1/2}}$ & 5--11 & 0.8628 & 0.880(6) & 0.798 \\
$b_{\phi\psi_{3/2}}$ & 5--11 & 0.6565 & 0.71(2) & 0.651 \\
$b_{S[\psi\psi]_S}$ & 5--11 & 0.8877 & 1.45(21) (n.c.) & -- \\
$b_{t[\psi\psi]_T}$ ($L_z=1$) & 5--11 & 0.4938 & 0.83(13) & -- \\
\midrule
$C_{\psi\psi[\psi\psi]_T^{L_z=1}}$ & 5--11 & 1.342 & 1.40(4) & 1.414 \\
$C_{\psi_{1/2}\psi_{-1/2}[\psi\psi]_T}$ & 5--11 & 0.729 & 1.14(17)  & 1.000 \\
$C_{\psi\bar\psi[\psi\psi]_S}$ & 5--11 & 1.020 & 1.01(5) & 1.000 \\
$C_{\psi\bar\psi[\psi\psi]_A}$ & 5--11 & 1.036 & 0.93(4) & 1.000 \\
$C_{\psi\bar\psi[\psi\psi]_T}$ & 5--11 & 1.049 & 0.98(3) & 1.000 \\
\bottomrule
\end{tabular}}
\end{table}
\begin{table}[t]
\centering\footnotesize
\caption{O(3) $R_z(\pi)$ defect: couplings, one-point functions, $C_D$ and defect-defect OPE coefficients.}
\label{tab:o3r_ope}
\fitwidth{%
\begin{tabular}{lcccc}
\toprule
quantity & sizes & largest & FS & free / GFF \\
\midrule
$b_{\phi\psi_{1/2}}$ & 5--11 & 0.8507 & 0.859(3) & 0.798 \\
$b_{\phi_3\hat\phi_3}$ & 5--11 & 1.0314 & 1.07(1) & 1.000 \\
$b_{\phi_3\hat\phi_3^{(1)}}$ & 5--11 & 0.7130 & 0.74(1) & 0.707 \\
$b_{S\hat\phi_3^2}$ & 5--11 & 0.5433 & 0.51(1) & -- \\
$b_{t_{33}\hat\phi_3^2}$ & 5--11 & 0.8778 & 1.00(5) & -- \\
$a_{t_{33}}$ & 5--11 & 0.1794 & 0.204(9) & 0.184\\
$b_{t_{33}\hat D}$ & 5--11 & 0.3917 & n.c. & -- \\
$C_D$ (Ward, $t_{33}$) & 5--11 & 0.1244 & n.c. & 0.101 \\
\midrule
$C_{\psi\psi(2,1)}$ & 5--11 & 1.351 & 1.41(3) & 1.414 \\
$C_{\psi_{1/2}\psi_{-1/2}(2,0)}$ & 5--11 & 0.737 & 1.14(16) & 1.000 \\
$C_{\hat\phi_3\hat\phi_3\hat\phi_3^2}$ & 5--11 & 1.078 & 1.14(7) & 1.414 \\
$C_{\psi\hat\phi_3[\psi\hat\phi_3]}$ & 5--11 & 0.941 & 0.88(2) & 1.000 \\
\bottomrule
\end{tabular}}
\end{table}
\begin{table}[t]
\centering\footnotesize
\caption{O(3) reflection defect: couplings, one-point functions, $C_D$ and defect-defect OPE coefficients. GFF value of $(\hat\phi^{(0)})^3$ in $\hat\phi\times(\hat\phi^{(0)})^2$: $\sqrt3$.}
\label{tab:o3z_ope}
\fitwidth{%
\begin{tabular}{lcccc}
\toprule
quantity & sizes & largest & FS & free / GFF \\
\midrule
$b_{\phi_3\psi^{(3)}_{1/2}}$ & 4--11 & 0.8395 & 0.841(3) & 0.798 \\
$b_{\phi_3\psi^{(3)}_{3/2}}$ & 4--11 & 0.6445 & 0.70(2) & 0.651 \\
$b_{\phi\hat\phi^{(0)}}$ & 4--11 & 1.0152 & 1.031(6) & 1.000 \\
$b_{\phi\hat\phi^{(1)}}$ & 4--11 & 0.7064 & 0.74(1) & 0.707 \\
$b_{S|\hat\phi|^2}$ & 4--11 & 0.7749 & 0.73(2) & -- \\
$b_{t_{33}|\hat\phi|^2}$ & 4--11 & 0.6467 & 0.75(4) & -- \\
$b_{t(\hat\phi^{(0)})^2}$ & 4--11 & 1.0197 & 1.07(2) & -- \\
$a_{t_{33}}$ & 4--11 & -0.1730 & -0.181(3) & -0.184\\
\midrule
$C_{\psi^{(3)}\psi^{(3)}\bar D}$ & 4--11 & 1.299 & 1.41(5) & 1.414 \\
$C_{\psi^{(3)}\bar\psi^{(3)}[\psi^{(3)}\psi^{(3)}]}$ & 4--11 & 1.018 & 0.96(3) & 1.000 \\
$C_{\hat\phi\hat\phi^\dagger|\hat\phi|^2}$ & 4--11 & 0.687 & 0.75(6) & 1.000 \\
$C_{\hat\phi\hat\phi^\dagger[\hat\phi\partial\hat\phi]_A}$ & 4--11 & 0.753 & 0.740(6) & $\sqrt{\hat\Delta_{\hat\phi}}\simeq0.70$ \\
$C_{\hat\phi\hat\phi(\hat\phi^{(0)})^2}$ & 4--11 & 1.218 & 1.28(5) & 1.414 \\
$C_{\hat\phi^{(0)}\hat\phi^{(1)}[\hat\phi^{(0)}\hat\phi^{(1)}]}$ & 4--11 & 0.904 & 1.05(7) & 1.000 \\
$C_{\hat\phi\psi^{(3)}[\hat\phi\psi^{(3)}]}$ & 4--11 & 0.906 & 0.98(5) & 1.000 \\
$C_{\hat\phi(\hat\phi^{(0)})^2(\hat\phi^{(0)})^3}$ & 4--11 & 1.347 & 1.52(12)  & 1.732 \\
$C_{\hat\phi^{(0)}\hat\phi^{(1)\dagger}\bar D}$ & 4--11 & 0.118 & 0.19(3) (n.c.)  & 0 (free) \\
\bottomrule
\end{tabular}}
\end{table}

\section{Finite-size analysis}
\label{app:fits}
This appendix summarizes how the    extrapolated estimates  quoted in the main text are obtained, and which observables are most
affected by finite-size corrections.

\paragraph{Extrapolation.} On the fuzzy sphere the natural expansion parameter of finite-size corrections is the
inverse radius $1/R\propto\Norb^{-1/2}$: corrections to defect dimensions from irrelevant defect operators scale as
$R^{-(\Dh-1)}$, those to matrix elements from descendant admixtures as $1/R$~\cite{Hu:2023xak,Hu:2023ghk}, and the exponents
of the leading corrections are not known a priori for most of our observables. We therefore fit each quantity with the
two two-parameter forms $X(\Norb)=X_\infty+c\,\Norb^{-1/2}$ and $X_\infty+c\,\Norb^{-1}$ on the four largest sizes; for dimensions
we do this in both normalizations. The quoted value is the centre of the resulting estimates and the quoted error is half
their spread, which is dominated by the difference between the two forms and is our estimate of the systematic
uncertainty of the extrapolation. Three-parameter forms ($X_\infty+c\,\Norb^{-1/2}+c'\Norb^{-1}$, $X_\infty+c\,\Norb^{-1}+c'\Norb^{-2}$) on the five
largest sizes are listed for representative quantities in Table~\ref{tab:fits}; they are stable for the Ising and O(2)
data sets and less stable for the shorter O(3) sequences. Quantities that still drift strongly (high transverse spins and
some couplings to operators made of two different modes) are marked as not converged (n.c.): their extrapolated values are shown for
orientation only and should not be read as reliable infinite-size limits. These forms    sample a limited family of  power-law corrections.    A flowing coupling can also produce crossover dependence; near a fixed point a small positive stability exponent instead gives a slow power law. The zero-mode mass flow~\cite{MCheng2026a} motivates checking both possibilities for  partial twists,    but does not establish logarithmic corrections at $d=3$. The quoted fit uncertainties do not cover all such alternatives or constitute a proof of the asymptotic limit .

\paragraph{Identification in near-degenerate sectors.} Several two-mode operators    share the same quantum numbers, so level order alone is not a reliable label. For  the reflection defect,    the twisted pair  $[\psi^{(3)}\psi^{(3)}]_{L_z=1}$ and  the appropriate neutral untwisted pair  $[\hat\phi^{(0)}\hat\phi^{(1)\dagger}]$    are degenerate in the free theory. By contrast, the neutral spin-zero pairs  $|\hat\phi|^2$ and $[\psi^{(3)}\bar\psi^{(3)}]$    have different free dimensions, although their interacting eigenstates can still mix . We identify such operators by the channel in which their OPE coefficient,
Eq.~\eqref{eq:C3}, is largest, restricted to the correct charge conjugation, and the displacement as the
symmetry-allowed level that approaches 2 in bulk normalization with increasing size and is continuously connected
across sizes. All identifications are stable over the full range of sizes.

\begin{figure}[t]
\centering
\includegraphics[width=\linewidth]{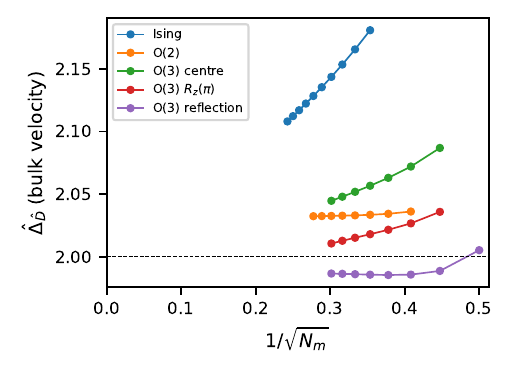}
\caption{Dimension of the displacement operator with the bulk velocity for the five $\mathbb Z_2$ defects. It approaches the exact value 2;
with the O($N$) model of Ref.~\cite{Guo:2025onf} it is within $4\%$ of 2 at all sizes and within $2\%$ at the largest sizes, while the Ising model of
Ref.~\cite{Zhu:2022gjc} has larger defect-localized corrections.}
\label{fig:disp}
\end{figure}

\paragraph{Displacement operator.} Since its dimension is fixed to $\Dh_{\hat D}=2$ by the Ward identity, the displacement operator
is the cleanest probe of the finite-size corrections localized at the defect. Figure~\ref{fig:disp} shows its dimension in the
bulk-velocity normalization. Its offset from 2    diagnoses deviations from the conformal spectrum, including core and bulk finite-size effects; for partial twists it can also receive crossover corrections . At
$\Norb=17$ the bulk-velocity estimate of the Ising displacement dimension is $2.108$, an absolute offset of about $0.11$, i.e.\
roughly $5\%$ relative to the exact value 2, decreasing monotonically from $9\%$ at $\Norb=8$, whereas the corresponding
relative deviation is at most $4\%$ at all sizes and at most $2\%$ at the largest sizes for the O($N$) model of
Ref.~\cite{Guo:2025onf}, whose densities are closer to the continuum fields. The difference between the bulk-velocity and
displacement normalizations is therefore included in the uncertainty of all extrapolated dimensions.

\paragraph{Convergence of the observables.} High transverse spins, $C_D$, and the couplings to operators made of two different
modes are the slowest observables; the bulk-to-defect couplings of the lowest modes and the OPE coefficients of non-interacting
channels are the fastest. For most quantities the $1/\sqrt{\Norb}$ and $1/\Norb$ extrapolations differ by $1$--$3\%$, which is the
dominant systematic uncertainty of this work (Table~\ref{tab:fits}).

\paragraph{Dependence on the fit form.} Table~\ref{tab:fits} lists, for representative quantities, the infinite-size values obtained
with the four finite-size forms. The two two-parameter forms in $\Norb^{-1/2}$ and $\Norb^{-1}$ typically differ by $1$--$3\%$. For the
defect dimensions of the Ising ($\Norb\le17$) and O(2) ($\Norb\le13$) data sets the three-parameter fits lie between them, whereas for
the couplings and OPE coefficients they scatter more widely; for the shorter O(3) sequences ($\Norb\le11$) the three-parameter fits are
less stable. They are therefore not used for the adopted values, which are the centre and half-spread of the two two-parameter fits,
in both normalizations for dimensions. For the untwisted $s=0$ modes all four forms give values below $\frac{1}{2}$.

\begin{table*}[t]
\centering\footnotesize
\caption{Stability of the extrapolations for representative quantities. Columns 4--7: infinite-size values from the fits $X_\infty+c/\sqrt{N_m}$ and $X_\infty+c/N_m$ (four largest sizes) and $X_\infty+c/\sqrt{N_m}+c'/N_m$ and $X_\infty+c/N_m+c'/N_m^2$ (five largest sizes). Adopted: centre and half-spread of the two two-parameter fits (for dimensions the adopted value in the main tables also includes the other normalisation). The three-parameter fits are less stable for the shorter O(3) data sets ($N_m\le11$).}
\label{tab:fits}
\fitwidth{%
\begin{tabular}{llccccccc}
\toprule
quantity & sizes & largest & $1/\sqrt{N_m}$ & $1/N_m$ & $1/\sqrt{N_m}+1/N_m$ & $1/N_m+1/N_m^2$ & adopted & reference\\
\midrule
Ising $\hat\Delta_{\psi_{1/2}}$ ($\hat D$ norm.) & 8--17 & 0.9192 & 0.908 & 0.914 & 0.910 & 0.913 & 0.911(3) & MC 0.9187(6)\\
Ising $\hat\Delta_{\psi_{1/2}}$ (bulk vel.) & 8--17 & 0.9688 & 0.893 & 0.933 & 0.918 & 0.929 & 0.91(2) & \\
\specialrule{0.02em}{0.25ex}{0.25ex}
Ising $\hat\Delta_{s}$ ($\hat D$ norm.) & 8--17 & 2.2688 & 2.315 & 2.291 & 2.294 & 2.292 & 2.30(1) & MC 2.27(1)\\
Ising $\hat\Delta_{s}$ (bulk vel.) & 8--17 & 2.3912 & 2.282 & 2.340 & 2.308 & 2.332 & 2.31(3) & \\
\specialrule{0.02em}{0.25ex}{0.25ex}
Ising $\hat\Delta_{p^o}$ ($\hat D$ norm.) & 8--17 & 2.9206 & 2.704 & 2.818 & 2.731 & 2.797 & 2.76(6) & MC 2.9(2)\\
Ising $\hat\Delta_{p^o}$ (bulk vel.) & 8--17 & 3.0783 & 2.645 & 2.872 & 2.759 & 2.845 & 2.76(11) & \\
\specialrule{0.02em}{0.25ex}{0.25ex}
Ising $b_{\sigma\psi_{1/2}}$ & 8--17 & 0.8613 & 0.856 & 0.859 & 0.844 & 0.855 & 0.857(3) & MC 0.968(2)\\
\specialrule{0.02em}{0.25ex}{0.25ex}
Ising $a_\epsilon$ (raw) & 8--17 & -0.2064 & -0.146 & -0.178 & -0.091 & -0.157 & -0.16(2) & MC $-0.167(4)$\\
\specialrule{0.02em}{0.25ex}{0.25ex}
Ising $\lambda_{\sigma\sigma\epsilon}$ ($T$-corr.) & 8--17 & 1.0943 & 0.999 & 1.049 & 1.071 & 1.054 & 1.02(3) & boot.\ 1.0519\\
\specialrule{0.02em}{0.25ex}{0.25ex}
Ising $C_{\psi\psi\bar D}$ & 8--17 & 1.3025 & 1.357 & 1.328 & 1.306 & 1.323 & 1.34(1) & boot.\ 1.345\\
\specialrule{0.02em}{0.25ex}{0.25ex}
Ising $C_{\psi\bar\psi s}$ & 8--17 & 1.0229 & 0.933 & 0.980 & 0.817 & 0.942 & 0.96(2) & boot.\ 0.95\\
\specialrule{0.02em}{0.25ex}{0.25ex}
Ising $C_{\psi\bar\psi p^o}$ & 8--17 & 0.8822 & 1.005 & 0.941 & 0.949 & 0.943 & 0.97(3) & boot.\ 0.988\\
\midrule
O(2) $\hat\Delta_{\Psi_{1/2}}$ ($\hat D$ norm.) & 6--13 & 0.9068 & 0.879 & 0.894 & 0.888 & 0.893 & 0.886(7) & one loop 0.900\\
O(2) $\hat\Delta_{\Psi_{1/2}}$ (bulk vel.) & 6--13 & 0.9215 & 0.892 & 0.908 & 0.905 & 0.907 & 0.900(8) & \\
\specialrule{0.02em}{0.25ex}{0.25ex}
O(2) $\hat\Delta_{\mathcal S}$ ($\hat D$ norm.) & 6--13 & 2.3320 & 2.418 & 2.372 & 2.378 & 2.373 & 2.39(2) & one loop 2.20\\
O(2) $\hat\Delta_{\mathcal S}$ (bulk vel.) & 6--13 & 2.3698 & 2.453 & 2.409 & 2.422 & 2.412 & 2.43(2) & \\
\midrule
O(3) centre $\hat\Delta_{\psi_{1/2}}$ ($\hat D$ norm.) & 5--11 & 0.9015 & 0.846 & 0.876 & 0.832 & 0.866 & 0.86(2) & one loop 0.886\\
O(3) centre $\hat\Delta_{\psi_{1/2}}$ (bulk vel.) & 5--11 & 0.9217 & 0.833 & 0.881 & 0.837 & 0.871 & 0.86(2) & \\
\specialrule{0.02em}{0.25ex}{0.25ex}
O(3) $R_z(\pi)$ $\hat\Delta_{\hat\phi_3}$ ($\hat D$ norm.) & 5--11 & 0.4940 & 0.448 & 0.473 & 0.426 & 0.462 & 0.46(1) & free value 0.5\\
O(3) $R_z(\pi)$ $\hat\Delta_{\hat\phi_3}$ (bulk vel.) & 5--11 & 0.4966 & 0.440 & 0.471 & 0.419 & 0.459 & 0.46(2) & \\
\specialrule{0.02em}{0.25ex}{0.25ex}
O(3) $R_z(\pi)$ $\hat\Delta_{\hat\phi_3^2}$ ($\hat D$ norm.) & 5--11 & 1.3367 & 1.242 & 1.293 & 1.196 & 1.271 & 1.27(3) & \\
O(3) $R_z(\pi)$ $\hat\Delta_{\hat\phi_3^2}$ (bulk vel.) & 5--11 & 1.3439 & 1.220 & 1.287 & 1.175 & 1.262 & 1.25(3) & \\
\specialrule{0.02em}{0.25ex}{0.25ex}
O(3) refl.\ $\hat\Delta_{\hat\phi^{(0)}}$ ($\hat D$ norm.) & 4--11 & 0.5070 & 0.481 & 0.495 & 0.471 & 0.490 & 0.488(7) & free value 0.5\\
O(3) refl.\ $\hat\Delta_{\hat\phi^{(0)}}$ (bulk vel.) & 4--11 & 0.5037 & 0.479 & 0.492 & 0.471 & 0.487 & 0.486(7) & \\
\specialrule{0.02em}{0.25ex}{0.25ex}
O(3) refl.\ $\hat\Delta_{|\hat\phi|^2}$ ($\hat D$ norm.) & 4--11 & 1.4598 & 1.453 & 1.457 & 1.397 & 1.443 & 1.455(3) & \\
O(3) refl.\ $\hat\Delta_{|\hat\phi|^2}$ (bulk vel.) & 4--11 & 1.4501 & 1.447 & 1.449 & 1.394 & 1.437 & 1.448(3) & \\
\bottomrule
\end{tabular}}
\end{table*}
\section{Plateau of the defect-channel ratio}
\label{app:plateau}
\begin{figure}[t]
\centering
\includegraphics[width=0.9\linewidth]{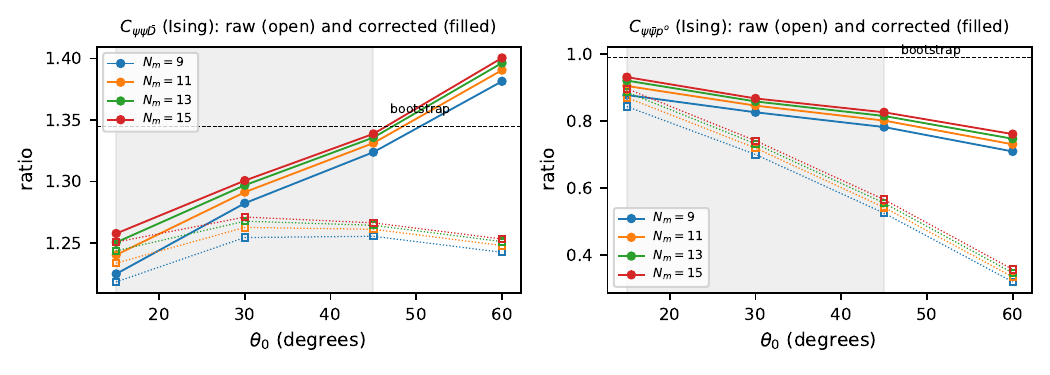}
\caption{The defect-channel ratio of Eq.~\eqref{eq:C3} as a function of $\theta_0$ for a well-behaved coefficient (left,
$C_{\psi\psi\bar D}$) and for the coefficient without a true plateau (right, $C_{\psi\bar\psi p^o}$), at four sizes. Open
squares: raw ratio; filled circles: after the kinematic factor $(\cos\theta_0)^{-(\Dh_1-\Dh_2-\Dh_3)}$. The shaded band is the
window $\theta_0\in[15^\circ,45^\circ]$ used for the estimate; the dashed line is the bootstrap value~\cite{Gaiotto:2013nva}.}
\label{fig:plateau}
\end{figure}

Figure~\ref{fig:plateau} illustrates the ratio method of Eq.~\eqref{eq:C3} for two defect-defect coefficients of the Ising twist
defect. For $C_{\psi\psi\bar D}$ the raw ratio (open squares) varies with $\theta_0$ mainly through the kinematics of the defect channel;
after the factor $(\cos\theta_0)^{-(\Dh_1-\Dh_2-\Dh_3)}$ (filled circles) it changes by only a few per cent over the window
$\theta_0\in[15^\circ,45^\circ]$ and approaches the bootstrap value $1.345$ as $\Norb$ grows. For $C_{\psi\bar\psi p^o}$ the corrected
ratio decreases monotonically across the window, from $0.93$ to $0.72$ at the largest size, so that no true plateau forms. We
attribute this to the next $\mathbb Z_2$-odd defect operators in the channel, which are not far above $\psi_{1/2}$ and produce
corrections of relative order $\theta_0^{\Dh_2'-\Dh_2}$; the quoted error of $C_{\psi\bar\psi p^o}$, which contains the half-spread over
the window, should therefore be read as a lower bound. Outside the window the method fails for two reasons: for
$\theta_0\lesssim\ell_B/R$ the    continuum defect expansion is contaminated by core-scale orbital effects , and at large $\theta_0$
the descendant corrections of order $\theta_0^2$ grow. This is why the estimate is restricted to $15^\circ$--$45^\circ$.

\bibliography{refs}

\end{document}